\documentclass[superscriptaddress,groupedaddress,nofootnoteinbib,11pt]{article}
\pdfoutput=1 % if your are submitting a pdflatex (i.e. if you have images in pdf, png or jpg format)
\usepackage{jheppub}

\usepackage{amsmath, amsfonts, amsthm, amssymb, graphicx, color, hyperref}
\usepackage{subfigure, array}
\usepackage[utf8]{inputenc}
\usepackage[normalem]{ulem}

\allowdisplaybreaks[3]
\makeatletter
\newcommand{\thickhline}{%
    \noalign {\ifnum 0=`}\fi \hrule height 1pt
    \futurelet \reserved@a \@xhline
}
\newcolumntype{"}{@{\hskip\tabcolsep\vrule width 1pt\hskip\tabcolsep}}
\makeatother
\def \bal#1\eal  {\begin{align} #1 \end{align}}
\def\({\left(}
\def\){\right)}
\def\[{\left[}
\def\]{\right]}
\def\<{\left\langle}
\def\>{\right\rangle}
\def\dla{\<\!\!\!\<}
\def\dra{\>\!\!\!\>}
\def\d{\mathrm{d}}

\newcommand{\eref}[1]{Eq~(\ref{#1})}
\newcommand{\f}[2]{\frac{#1}{#2}}
\newcommand{\bim} {\begin{itemize}[noitemsep]}

\newcommand{\eim}{\end{itemize}}
\newcommand{\be} {\begin{equation}}
\newcommand{\ee} {\end{equation}}
\newcommand{\bc}{\begin{center}}
\newcommand{\ec}{\end{center}}
\newcommand{\lt}{\left}
\newcommand{\rt}{\right}
\newcommand{\nn} {\nonumber\\}

\newcommand{\marrow}{~~\Longrightarrow~~}

\newcommand{\im}{{\rm Im}}

\newcommand{\pd} {\partial}
\newcommand{\mc} {\mathcal}

\newcommand{\ai}{{\alpha}}

\newcommand{\gi}{{\gamma}}

\newcommand{\ri}{{\rho}}
\newcommand{\si}{{\sigma}}
\newcommand{\li}{{\lambda}}

\newcommand{\ei}{{\eta}}

\newcommand{\epi}{\epsilon}
\newcommand{\thi}{\theta}
\newcommand{\Gi}{\Gamma}
\newcommand{\Li}{\Lambda}
\def \= {\equiv}
\newcommand{\beq} {\begin{equation}}
\newcommand{\eeq} {\end{equation}}
\newcommand{\mtL} {{ L}}
\newcommand{\mtM} {{ M}}
\newcommand{\mtP} {{ P}}
\newcommand{\mtQ} {{ Q}}
\newcommand{\mtW} {{ W}}
\newcommand{\ki}{\kappa}
\title{New positivity bounds from full crossing symmetry}
\author[a,b]{Andrew J. Tolley,}
\author[c]{Zi-Yue Wang}
\author[d,e]{and Shuang-Yong Zhou}
\affiliation[a]{Theoretical Physics, Blackett Laboratory, Imperial College, London, SW7 2AZ, U.K.}
\affiliation[b]{CERCA, Department of Physics, Case Western Reserve University, 10900 Euclid Ave, Cleveland, OH 44106, USA}
\affiliation[c]{School of Gifted Young, University of Science and Technology of China, Hefei, Anhui 230026, China}
\affiliation[d]{Interdisciplinary Center for Theoretical Study, University of Science and Technology of China, Hefei, Anhui 230026, China}
\affiliation[e]{Peng Huanwu Center for Fundamental Theory, Hefei, Anhui 230026, China}

\emailAdd{a.tolley@imperial.ac.uk}
\emailAdd{metrictensor@mail.ustc.edu.cn}
\emailAdd{zhoushy@ustc.edu.cn}

\preprint{\small USTC-ICTS/PCFT-20-38}

\date{\today}

\abstract{
Positivity bounds are powerful tools to constrain effective field theories. Utilizing the partial wave expansion in the dispersion relation and the full crossing symmetry of the scattering amplitude, we derive several sets of generically nonlinear positivity bounds for a generic scalar effective field theory: We refer to these as the $PQ$, $D^{\rm su}$, $D^{\rm  stu}$ and $\bar{D}^{\rm stu}$ bounds. While the $PQ$ bounds and $D^{\rm su}$ bounds only make use of the $s\leftrightarrow u$ dispersion relation, the $D^{\rm  stu}$ and $\bar{D}^{\rm stu}$ bounds are obtained by further imposing the $s\leftrightarrow t$ crossing symmetry. In contradistinction to the linear positivity for scalars, these inequalities can be applied to put upper and lower bounds on Wilson coefficients, and are much more constraining as shown in the lowest orders. In particular we are able to exclude theories with soft amplitude behaviour such as weakly broken Galileon theories from admitting a standard UV completion. We also apply these bounds to chiral perturbation theory and we find these bounds are stronger than the previous bounds in constraining its Wilson coefficients.
}

\begin{document}
\maketitle
\flushbottom

\section{Introduction}

Effective field theories (EFTs) are widely used in modern physics. It is a general framework that reflects the fact that physics at different energy scales are largely decoupled as long as there is a healthy hierarchy between the scales. The standard practice to construct an EFT is to write down all possible operators that are consistent with the symmetries of the problem, be it gauge or global, and supply them with arbitrary Wilson coefficients. These coefficients should be fixed or constrained by using the experimental data or matching to the UV theory, and then the EFT can be used to predict new physical phenomena. This is of course a foolproof procedure. However, quite often, experimental data are scarce and imprecise, and the UV theory is unknown or very difficult to be matched to. After all, those are often the reasons why EFT is used in the first place.

Positivity bounds are constraints on the scattering amplitude that one can derive by assuming the UV theory satisfies some of the most fundamental properties of physics, such as Lorentz invariance, unitarity, crossing symmetry, polynomial (or exponential) boundedness and, crucially, analyticity. The textbook application of unitarity is the optical theorem, which links the imaginary part of the amplitude in the forward limit to the total cross section. The positivity of the cross section then implies that the imaginary part of the amplitude is positive, which can be used to prove the forward limit positivity bounds \cite{Adams:2006sv} (see \cite{Pham:1985cr, Pennington:1994kc, Ananthanarayan:1994hf, Comellas:1995hq} for related earlier work). Since the total angular momentum is conserved in a scattering process, the amplitude can be decomposed into different partial waves, each of which inherits unitarity. Thanks to partial wave unitary and properties of the Legendre polynomials (Wigner's small $d^J_{ab}$ matrices for particles with spin), one can show that the $t$ (Mandelstam variable $s,t, u$) derivatives of the imaginary part of the amplitude is still positive. Also, unitarity, coupled with polynomial boundedness, allows one to derive the Froissart-Martin bound \cite{Froissart:1961ux, Martin:1962rt, Jin:1964zza}, which limits the amplitude to grow slower than $s^2$ at large $|s|$. This is useful because, with appropriate subtractions, it implies that a contour integral at large $|s|$ should vanish. Then, by Cauchy's integral formula in the complex $s$ plane, analyticity gives rise to a dispersion relation, which casts the amplitude in terms of an integral along the discontinuities along the real $s$ axis. The dispersion relation, together with the positivity of the $t$ derivative of the amplitude, can be used to derive an infinite number of generalized positivity bounds \cite{deRham:2017avq,deRham:2017zjm}, to be referred to as the $Y$ bounds in this paper. Thanks to the analyticity in the complex $t$ plane \cite{Martin:1965jj}, the positivity bounds can also be extended away from the forward limit \cite{deRham:2017avq} (see \cite{Pennington:1994kc,Vecchi:2007na,Manohar:2008tc,Nicolis:2009qm, Bellazzini:2016xrt} for earlier work).

These positivity bounds have useful applications in EFTs because at energies well below the cutoff the EFT scattering amplitudes can approximate the full amplitude very well. Therefore, we can replace the full amplitude in the positivity bounds with the amplitude computed from the EFT, which contains Wilson coefficients of the EFT, and so the positivity bounds become inequalities for the Wilson coefficients. Typically, the EFT parameter space spanned by the Wilson coefficients is very huge. For example, viewing the Standard Model as an EFT, in addition to the textbook renormalizable Lagrangian, it should be augmented with many more higher dimensional operators. Specifically, at dimension 6, 8 and 10, there are 84, 993 and 15456  independent operators respectively \cite{Henning:2015alf}. The positivity bounds, on the other hand, carve out the part of the parameter space that is consistent with the fundamental physical principles mentioned above. Indeed, they are often very effective in restricting the parameter space.  Recently, there has been a lot of interest in applying the positivity bounds in particle physics, gravitational theories and cosmology \cite{deRham:2018qqo, deRham:2017imi, Zhang:2018shp, Bi:2019phv, Zhang:2020jyn, Fuks:2020ujk, Yamashita:2020gtt, Remmen:2020vts, Remmen:2019cyz, Bellazzini:2018paj, Bellazzini:2015cra, Cheung:2016yqr,  Cheung:2016wjt, Bellazzini:2017fep, Bonifacio:2016wcb,  Bellazzini:2017bkb,  Bonifacio:2018vzv, Distler:2006if, Bellazzini:2019xts, Melville:2019wyy, deRham:2019ctd, Alberte:2019xfh, Alberte:2019zhd}. In particular, positivity bounds have been used to constrain Standard Model EFT \cite{Zhang:2018shp, Bi:2019phv, Zhang:2020jyn, Fuks:2020ujk, Yamashita:2020gtt, Remmen:2020vts, Remmen:2019cyz, Bellazzini:2018paj}. For example, it has been shown that the vast majority of the parameter space of the dimension 8 anomalous quartic gauge couplings is excluded by the positivity bounds \cite{Zhang:2018shp, Bi:2019phv, Yamashita:2020gtt, Remmen:2019cyz}. For another example, strong constraints can also be obtained for Horndeski theory when pairing together with cosmological parameter estimation analysis \cite{Melville:2019wyy}. This helps the experimental searches for new physics, as experimentalists can pay more attention to the parameter space consistent with the positivity bounds. On other hand, should signals indicate violations of the positivity bounds, it would imply that some of the cherished physical principles be violated at high energies \cite{Fuks:2020ujk}.

In this paper, we will demonstrate that there are a couple of new ways to extract more positivity bounds using the dispersion relation and crossing symmetry for the case of scalar field theories in $D$ dimensions. These fall into two classes which we refer to as the $PQ$ bounds and the $D$ bounds. To derive the $PQ$ bounds, we follow a strategy used in deriving the $Y$ bounds in \cite{deRham:2017avq} and use the relaxing inequality, that is, use the fact that a quantity gets relaxed (an inequality is obtained) when one fixes some factor of a positive integrand to the lower limit of the integration. Different from the $Y$ bounds case, here we relax a quantity in two directions. This allows us to nonlinearly combine the derivatives of the amplitude to obtain specific nonlinear positivity bounds. To derive the $D$ bounds, we expand the amplitude in the dispersion integral with partial waves and make use of the angular momentum dependence of the partial wave coefficients. In this case, one also relaxes the integrand of an integration, but now we relax the polynomial of the partial wave number $\ell$ to its minimum. The triple crossing symmetry can improve this procedure because the triple symmetry requires that the integration of some polynomials of $\ell$ be zero.

These new positivity bounds are very constraining. As we will show in Section \ref{sec:comp}, generically, the $Y$, $PQ$ and $D$ bounds are overlapping with each other, and combining them together can improve constraints on the parameter space by more than one order of magnitude in percentages. In particular, for the generic parameter space of the 2-to-2 scattering amplitude truncated up to order $E^{12}$ in an energy expansion, only about $0.2\%$ of the total parameter space is consistent with the fundamental principles of the scattering amplitude. The effectiveness of these new bounds can also be seen in their phenomenological usages. As a simple example, we apply the new positivity bounds to chiral perturbation theory (the theory that marks the beginning of applications of the modern EFT idea), and we will show that the new bounds improve the bounds on the $\bar l_1$ and $\bar l_2$ parameters in the theory, compared to the previous positivity bounds.

As another concrete example of applications of these new positivity bounds, we show that the new positivity bounds eliminate theories with soft behaviour for scattering amplitudes, i.e. for which the leading few terms in an energy expansion of the low energy amplitude are suppressed relative to naive expectations \cite{Cheung:2014dqa,Cheung:2016drk,Bellazzini:2016xrt,Hinterbichler:2014cwa}. One notable example is the case of theories with weakly broken Galileon symmetries. Galileon theory is a scalar field theory with an enhanced shift symmetry $\pi \to \pi + c+b_\mu x^\mu$ ($c,b_\mu$ being constant and $x^\mu$ being the spacetime coordinates) \cite{Nicolis:2008in} that arises in the decoupling limit of massive gravity theories \cite{Luty:2003vm,deRham:2010ik, deRham:2010kj} (see \cite{deRham:2014zqa} for a review). In this limit, one takes the (effective) graviton mass $m$ to zero and the Planck mass $M_P$ to infinity and keeps the strong coupling scale $\Li_3=(m^2M_P)^{1/3}$ fixed. As a result, Galileon theory captures the most important nonlinear deviations of massive gravity theories from general relativity.  Away from the decoupling limit, the degree of freedom represented by the Galileon is massive, and thus it is arguably motivated to add the mass term to Galileon theory. More generally, since the massless Galileon is ruled out by the linear positivity bounds \cite{Adams:2006sv}, it is generally regarded that a partial resolution is to weakly break the Galileon symmetry.
That is to say one may regard the fact that the Galileon violates positivity bounds as a marginal phenomena. A mass term is the most natural way to do so in the sense that the non-renormalization theorem of the Galileon theory still holds and all generated quantum corrections still enjoy the enhanced shift symmetry \cite{deRham:2017imi}, explicitly realised for example in the context of Galileon inflation \cite{Burrage:2010cu}. It was shown in \cite{deRham:2017imi} that there is still a surviving region in the parameter space where softly broken Galileon theory is compatible with the linear $Y$ positivity bounds. More generally in cosmological contexts it is natural to consider more general breaking terms \cite{Pirtskhalava:2015nla}.

In this paper, we will show that the new positivity bounds derived accounting for full (triple) crossing symmetry implies that the Minkowski spacetime field theory of a weakly broken Galileon symmetry is forbidden from admitting a standard UV completion. Similar arguments trivially extend to theories with higher order Galileon symmetries or higher order soft amplitude beahviour \cite{Cheung:2014dqa,Cheung:2016drk,Bellazzini:2016xrt,Hinterbichler:2014cwa}. The direct application of these results to theories of massive gravity is more subtle, as it is necessary to deal with the application of positivity bounds to particles with spin \cite{deRham:2017zjm} and the difficult issues related to massive and massless spin-2 $t$-channel poles \cite{Bellazzini:2019xts,Loges:2019jzs,Alberte:2020jsk,Tokuda:2020mlf}. These shall be considered elsewhere \cite{moreBoundsSpin}.

It should be stressed that while these results do clearly rule out Galileons and their extensions as the low energy limit of standard local field theories, they do not forbid their role as limits of more general gravitational type theories which admit weaker notion of locality \cite{Dvali:2012zc,Keltner:2015xda}. Indeed, the entire strength of the standard positivity bound story rests on the assumption that the scattering amplitude is bounded by $|s|^2$, at large $|s|$ and fixed momentum transfer, which is traditionally derived from the assumptions of polynomial or (linear) exponential boundedness. The validity of these assumptions in the gravitational context is unclear. In essense, since we typically do not expect local gauge invariant observables in a quantum theory of gravity, it is unclear why the scattering amplitude should respect locality in the usual manner. These issues are further closely intertwined with the technical issues in the applicability of positivity bounds in the presence of gravity \cite{Bellazzini:2019xts,Loges:2019jzs,Alberte:2020jsk,Tokuda:2020mlf}. \\

The paper is organized as follows: In Section \ref{sec:Ybounds}, we review the $Y$ positivity bounds derived in \cite{deRham:2017avq}, as we will compare the new bounds with the $Y$ bounds later, and also establish some notations along the way; In Section \ref{sec:newBoundExample}, as a warm-up, we derive some simple examples of the new positivity bounds; In Section \ref{sec:galileon}, we apply the first new positivity bounds to  theories with soft amplitudes - specifically the weakly broken Galileon theory and show that such soft amplitude theories cannot have an analytical UV completion; In Section \ref{sec:npbG}, we take a more systematical approach to derive a few sets of different positivity bounds, first using only the $s\leftrightarrow u$ symmetric dispersion and then further imposing the $s\leftrightarrow t$ symmetry; The best triple crossing symmetric bounds up to level $1/\mu^{10}$ in 4D are presented in explicit form; In Section \ref{sec:comp}, we explore the differences between the $Y$ bounds and the new positivity bounds; In Section \ref{sec:chpt}, we use these new bounds to constrain SU(2) chiral perturbation theory; We conclude in Section \ref{sec:sum}. \\

{\bf Note added:} While we were putting final touches on this draft, \cite{Bellazzini:2020cot} appeared which contains some overlap in results obtained through a slightly different method. In particular, these authors reach a similar conclusion about theories with soft amplitudes \cite{Cheung:2014dqa,Cheung:2016drk,Bellazzini:2016xrt,Hinterbichler:2014cwa}. In addition, after our paper appeared in arXiv,  \cite{Caron-Huot:2020cmc} and \cite{Guerrieri:2020bto} also arrived. The discussion of \cite{Caron-Huot:2020cmc} in particular is closely parallel to the present paper.

\section{Fixed-$t$ dispersion relations}
\label{sec:Ybounds}

In this section, {we shall introduce the fixed-$t$ dispersion relations that are the basis needed to derive the positivity bounds in the following sections. We shall focus on the case of a single scalar field.}
%we shall review the linear positivity bounds derived in \cite{deRham:2017avq} for the case of a single scalar, which can be conveniently formulated as a recurrence relation that defines positive $Y^{(2N,M)}$ quantities, which in turn are sums of derivatives of the scattering amplitude with respect to the Mandelstam variables. Slightly different from \cite{deRham:2017avq}, here we shall present the results in $D$ dimensions.
The formulas in the following are valid only, strictly speaking, for $D\geq 4$, but as we will see in Appendix \ref{sec:Dequto3}, with some appropriate definitions, {it is possible to include the $D=3$ case}.

The 2-to-2 scattering amplitude for scalar particles is a Lorentz invariant function of Mandelstam variables $s$, $t$ and $u$ that satisfy the constraint $s+t+u = 4m^2$ and the scattering angle $\thi$ can be expressed as
\be
\cos\thi = 1+\frac{2t}{s-4m^2} .
\ee
Choosing $s$ and $t$ as the independent variables, the amplitude $A(s,t)$ can be viewed as an analytic function with complex variables $s$ and $t$, except for certain poles and branch cuts already seen in perturbation theory. The partial wave expansion in $D$ dimensions is facilitated by $D$-dimensional generalization of the Legendre polynomials --- the Gegenbauer polynomials $C^{(\ai)}_{\ell}(x)$:
\bal
A(s,t)
\label{pwexp}
&=F(\ai)  \frac{{s}^{1/2}}{(s-4m^2)^{\ai}}  \sum_{\ell=0}^\infty (2\ell+2\ai)C^{(\ai)}_{\ell}(\cos\thi) a_\ell(s) , ~~~\ai = \frac{D-3}2   ,
\eal
where $F(\ai)=2^{4\ai+2}\pi^\ai \Gamma(\ai)$ is positive for $D\geq 4$. See Appendix \ref{sec:gegen} for a brief introduction of the Gegenbauer polynomials. In particular, the derivatives of the Gegenbauer polynomials have positive properties when evaluated at $x=1$:
\be
\left. \frac{\d^n}{\d x^n}C^{(\ai)}_{\ell}(x)\right|_{x=1} = \frac{ 2^{-2 \ai -n+1}\sqrt{\pi } \Gamma (\ell+n+2 \ai )}{ \Gamma (\ell-n+1) \Gamma (\ai )\Gamma \left(n+\ai +\frac{1}{2}\right)}\geq 0,~~~ n\geq 0   ,
\ee
and, also, $C^{(\ai)}_{\ell}(-x)=(-1)^\ell C^{(\ai)}_{\ell}(x)$. The $t\leftrightarrow u$ crossing symmetry of the amplitude $A(s,t)=A(s,u)$ implies that $\sum_{\ell=0}^\infty (2\ell+2\ai)C^{(\ai)}_{\ell}(1+\frac{2t}{s-4m^2}) a_\ell(s)=\sum_{\ell=0}^\infty (2\ell+2\ai)C^{(\ai)}_{\ell}(1+\frac{2u}{s-4m^2}) a_\ell(s)$, the right hand side of which can be written as $\sum_{\ell=0}^\infty (2\ell+2\ai)(-1)^\ell C^{(\ai)}_{\ell}(1+\frac{2t}{s-4m^2}) a_\ell(s)$, so we must have $a_{\ell={\rm odd}}(s)=0$. On the other hand, partial wave unitarity tells us that
\be
\label{pwuni}
|a_\ell(s)|^2\leq \im a_\ell(s), \quad s \geq 4 m^{2}  ,
\ee
so $\im a_\ell(s)$ is positive in the physical region $s \geq 4 m^{2}$. Combining these positive properties, we can infer that in the forward limit $t=\thi=0$ we have
\be
\label{Ast0positive}
\f{\d^n }{\d t^n}\im A(s,t=0) >0 ,   \quad s \geq 4 m^{2} ,~~~ n\geq 0  .
\ee

Although stronger analyticity may be assumed, a weaker analyticity condition, proved by Martin \cite{Martin:1965jj} from more basic assumptions, is already powerful in many applications, which states that for fixed $s$, $A(s,t)$ is analytic in the disk $|t|<4m^2$ modulo possible poles and for fixed $t$, $A(s,t)$ is analytic in the plane of $s$ except for possible poles and the branch cuts at $s>4m^2$ and $s<-t$. Fixed-$s$ analyticity allows us to Taylor expand $\im A(s,t)$ around $t=0$ in the disk $|t|<4m^2$ and the positivity of Eq \ref{Ast0positive} then implies
\be
\label{dndtImA}
\f{\d^n }{\d t^n}\im A(s,t) >0, \quad  \quad s \geq 4 m^{2}, ~~ 0 \leq t<4 m^{2}     .
\ee
On the other hand, the Jin-Martin extension of the Froissart-Martin \cite{Froissart:1961ux, Martin:1962rt,Jin:1964zza} (assumed to be valid in $D$ dimensions) implies that
\begin{equation}
\lim _{s \rightarrow \infty}|A(s, t)|<C s^{1+\varepsilon(t)}, \quad \varepsilon(t)<1  ,  \quad 0 \leq t<4 m^{2}   .
\end{equation}
Therefore, utilizing fixed-$t$ analyticity in the $s$ complex plane and Cauchy's integral formula,
we can derive a twice-subtracted dispersion relation (see e.g.~\cite{deRham:2017avq} for details)
\bal
A(s,t) &= a(t)  + \frac{\li}{m^2 -s} + \frac{\li}{m^2 -t} + \frac{\li}{m^2 -u}
\nn
&~~~~+  \int^{\infty}_{4m^2} \frac{\d \mu}{\pi (\mu-\mu_p)^2} \[ \frac{(s-\mu_p)^2}{\mu-s}+\frac{(u-\mu_p)^2}{\mu-u}   \]   \im A(\mu,t)    ,
\eal
where $\li$ is a constant, $\mu_p$ is a subtraction point that we can choose and $a(t)$ is an unknown function of $t$. With the pole contribution also subtracted, we can define
\bal
\label{orignalBrel}
B(s,t)&=A(s,t)- \frac{\li}{m^2 -s} - \frac{\li}{m^2 -t}- \frac{\li}{m^2 -u}
\\
\label{B0disrel}
& = a(t) +  \int^{\infty}_{4m^2} \frac{\d \mu}{\pi (\mu-\mu_p)^2} \[ \frac{(s-\mu_p)^2}{\mu-s}+\frac{(u-\mu_p)^2}{\mu-u}   \]   \im A(\mu,t)   .
\eal
{With these ingredients, particularly \eref{dndtImA}, we can derive an infinite number of positivity bounds with $s$ and $t$ derivatives ---  the $Y$ positivity bounds; see \cite{deRham:2017avq} for the recurrence relations for the positive $Y^{(2N,M)}$ quantities. As we will see in the following sections, the use of \eref{dndtImA} does not give rise to the optimal bounds. Indeed, better bounds can be extracted by using the positivity of $\im a_\ell(s)$ and the detailed properties of the Gegnebauer polynomials. To show this, in Section \ref{sec:comp}, we shall compare our new bounds against the $Y$ bounds.}

%To derive the $Y^{(2N,M)}$ positivity bounds, it is convenient to choose $\mu_p=-\bar t/2$ and make a change of variable from $s$ to $v=\bar s+\bar{t}/2$ so that we have
%\be
%\label{tildeBdr}
%\tilde{B}(v, t)=a(t)+\int_{4m^2}^{\infty} \frac{d \mu}{\pi(\bar{\mu}+\bar{t} / 2)} \frac{2 v^{2} \operatorname{Im} A(\mu, t)}{(\bar{\mu}+\bar{t} / 2)^{2}-v^{2}}   ,
%\ee
%where we have defined
%\be
%\bar s= s- \f43 m^2,~~ \bar t=t- \f43 m^2, ~~\bar u=u- \f43 m^2,
%\ee
%Also defining
%\be
%B^{(2 N, M)}(t)=\left.\frac{1}{M !} \partial_{v}^{2 N} \partial_{t}^{M} \tilde{B}(v, t)\right|_{v=0}  ,
%\ee
%we can derive the $Y$ positivity bounds \cite{deRham:2017avq}
%\bal
%Y^{(2 N, M)}(t) &\sout{=\sum_{r=0}^{\lfloor M / 2\rfloor} c_{r} B^{(2 N+2 r, M-2 r)}(t)   }
%\nn
%\label{Yrel}
%&~~~~ +\frac{1}{\mathcal{M}^{2}} \sum_{k \text { even }}^{\lfloor (M-1) / 2\rfloor}(2(N+k)+1) \beta_{k} Y^{(2(N+k), M-2 k-1)}(t)>0  ,
%\eal
%where $\lfloor\; \rfloor$ means taking the flooring integer, $\mc{M}^2=\f{t}2 +2m^2$ and $c_r$ and $\beta_k$ are the Taylor expansion coefficients of
%\be
%\operatorname{sech} \( \f{z} 2 \)=\sum_{k=0}^{\infty} c_{k} z^{2 k} \quad \text { and } \quad \tan \(\f{z} 2\)=\sum_{k=0}^{\infty} \beta_{k} z^{2 k+1} .
%\ee

Up to this point, we have only used the UV full amplitude to derive {the dispersion relation}. The reason why EFT comes into play is that at low energies the EFT amplitude approximates the full amplitude very well, to a desired order in the EFT power counting, so the bounds can be interpreted as constraints on the EFT. If one parametrizes the amplitude as
\be
\label{adef}
B(s,t) = \sum_{i,j=0}^{\infty} a_{i,j} x^i y^j = \sum_{i,j=0}^{\infty} \frac{\tilde{a}_{i,j}}{\Lambda^{4i+6j}} x^i y^j   ,
\ee
where $x$ and $y$ are triple crossing symmetric variables defined as
\be
x=-(\bar s\bar t + \bar s \bar u+\bar t \bar u) = \bar s^2 + \bar s \bar t + \bar t^2   ,
~~~~~~
y=-\bar s\bar t\bar u = \bar s^2 \bar t + \bar s \bar t^2    ,
\ee
with
\be
\bar s= s- \f43 m^2,~~ \bar t=t- \f43 m^2, ~~\bar u=u- \f43 m^2,
\ee
then one can express the positivity bounds as inequalities on the expansion coefficients $a_{i,j}$. In EFTs, they are directly linked to the Wilson coefficients. So the positivity bounds are constraints on the Wilson coefficients.

The lower limit of the integration in \eref{B0disrel} is from $4m^2$. This generally renders higher order $M$ applications of the $Y$ bounds contentless as they will be dominated by the terms with the largest powers of the inverse of the small (in comparison to $\Lambda$) mass $m$. However, since we can actually compute the imaginary part of the amplitude to a desired order within the EFT framework from $4m^2$ to $(\epi \Lambda)^2$, where $\Lambda$ is the cutoff and $\epi\lesssim 1$, so we can also subtract out the low energy part of the integral and define
\bal
\label{Bdisrel}
 B_{\epi\Lambda}(s,t) &= A(s,t)- \frac{\li}{m^2 -s}  - \frac{\li}{m^2 -t} - \frac{\li}{m^2 -u}
 \nn
 &~~~~~ -  \int^{(\epi\Lambda)^2}_{4m^2} \frac{\d \mu}{\pi (\mu-\mu_p)^2} \[ \frac{(s-\mu_p)^2}{\mu-s}+\frac{(u-\mu_p)^2}{\mu-u}   \]   \im A(\mu,t)
\\
\label{Bintegrand}
& = a(t) +  \int^\infty_{(\epi\Lambda)^2}  \frac{\d \mu}{\pi (\mu-\mu_p)^2} \[ \frac{(s-\mu_p)^2}{\mu-s}+\frac{(u-\mu_p)^2}{\mu-u}   \]   \im A(\mu,t)     .
\eal
With this, going through the same steps, we can derive improved $Y$ positivity bounds $Y_{\epi\Lambda}^{(2 N, M)}>0$ \cite{deRham:2017xox,deRham:2017imi}.
%\bal
%Y_{\epi\Lambda}^{(2 N, M)}(t) &=\sum_{r=0}^{\lfloor M / 2\rfloor} c_{r} B_{\epi\Lambda}^{(2 N+2 r, M-2 r)}(t)
%\nn
%\label{Yrelimp}
%&~~~~ +\frac{1}{\mathcal{M}^{2}} \sum_{k \text { even }}^{\lfloor (M-1) / 2 \rfloor}(2(N+k)+1) \beta_{k} Y_{\epi\Lambda}^{(2(N+k), M-2 k-1)}(t)>0   ,
%\eal
%where now $\mc{M}^2=(\epi\Lambda)^2+\f{t}2 -2m^2 \simeq (\epi\Lambda)^2$. From \eref{Yrelimp}, we see that
As the higher $t$-derivative $Y$ bounds are constructed by linearly combining derivatives of the amplitude with the lower $t$-derivative bounds, {a greater $\epi\Lambda$} will
%suppress the $t$-derivative bounds and hence
enhance the importance of the higher $t$-derivative bounds, in addition to the fact that the subtraction from ${4m^2}$ to $(\epi\Lambda)^2$ already improves the $Y$ bounds.

An often considered case is that the UV completion is weakly coupled and this weak coupling is also accessible at low energies. In this case, loop diagrams can be suppressed with respect to the tree diagrams by the UV weak coupling and the tree level amplitude already unitarizes  the amplitude in the UV, so we can have a tree level dispersion relation $B_{\rm tr}(s,t)$, which is similar to Eq (\ref{Bdisrel}) but the integrand is replaced with the tree level amplitude and the integration starts from $\Lambda_{\rm th}$, the energy scale of the first state that lies outside the EFT \cite{deRham:2017avq}. Then, we can similarly derive the tree level positivity bounds $Y_{\rm tr}^{(2 N, M)}$. Note that for a tree level amplitude, its imaginary part vanishes, so the $\epi\Lambda$ subtracted amplitude $B_{\epi\Lambda}(s,t)$ is the same as $B_{\rm tr}(s,t)$, so $Y_{\rm tr}^{(2 N, M)}$ is a special case of $Y_{\epi\Lambda}^{(2 N, M)}$.

%Note that, roughly speaking, the quantities $Y^{(2 N, M)}$ or $Y_{\epi\Lambda}^{(2 N, M)}$ are linear combinations of the $s$ and $t$ derivatives of the scattering amplitude. In contrast, with additional inputs from the partial wave expansion and crossing symmetry, the positivity bounds we will derive in the following are often nonlinear in the amplitude (and its $s$ and $t$ derivatives).

\section{New positivity bounds: Simple examples}
\label{sec:newBoundExample}

In this section, we will make further use of the dispersion relation (\ref{Bintegrand}) to extract some new positivity bounds. In deriving the  $Y$ positivity bounds, we essentially used the fact that the imaginary part of the amplitude is positive in appropriate ranges of $s$ and $t$, {\it i.e.}, \eref{dndtImA}. However, the partial wave expansion and partial wave unitarity actually contain more information, yet to be profited to derive new positivity bounds. Also, the dispersion relation (\ref{B0disrel}) or (\ref{Bintegrand}) are only manifestly $s\leftrightarrow u$ crossing symmetric, while the amplitude is actually triple crossing symmetric, which has not been used to derive the $Y$ positivity bounds. In this section we will take advantages of these new pieces of information to derive the first examples of new positivity bounds before taking a more systematical approach in Section \ref{sec:npbG}.

First, since the integrand of \eref{Bintegrand} is positive in the physical region $\mu>4m^2$, we can introduce a positive ``density distribution''
\be
\ri_{\ell,\ai}(\mu)=\frac{F(\ai)}{(\mu-\mu_p)^3}   \frac{\mu^{1/2}}{({\mu-4m^2})^{\ai}} (2\ell+2\ai)\im a_\ell(\mu) C_\ell^{(\ai)}(1)  ,
\ee
with
\be
\label{Cl1}
C^{(\ai)}_{\ell}(1) = \frac{\Gamma (\ell+D-3 )}{\Gamma (D-3 ) \Gamma (\ell+1)} =  \binom{\ell+D-4}{\ell} >0 ,~~~D\geq 4    ,
\ee
where $\binom{n}{k}=n!/[k!(n-k)!]$ are the binomial coefficients.  Then the dispersion relation can be written as
\be
\label{Bdisrel21}
B_{\epi\Lambda}(s,t) = a(t) +  \sum_\ell \int \d \mu \[ \frac{(s-\mu_p)^2}{\mu-s}+\frac{(u-\mu_p)^2}{\mu-u}   \] \frac{(\mu-\mu_p)\ri_{\ell,\ai}(\mu)}{C_\ell^{(\ai)}(1)} C^{(\ai)}_{\ell}\(1+\frac{2t}{\mu-4m^2}\)    ,
\ee
where for simplicity we have suppressed the summation and the integration limits, which are from $(\epi\Lambda)^2$ from $\infty$.

New positivity bounds are easiest to see when the derivatives of the amplitude are evaluated at $ s= t=0$ and the limit $(\epi\Lambda)^2\gg  m^2\to 0$ is taken for the expansion coefficients, which is the approach we take in this section. In other words, we shall evaluate $s$ and $t$ derivatives of $B_{\epi\Lambda}(s,t)$ at $s=t=0$, which leads to a dispersion relation where the integrand is a function of $\mu$ and $m^2$, and since the low limit of $\mu$ is $(\epi\Lambda)^2$, we can neglect all the subleading terms with $m^2$. Clearly, the $m\to0$ limit can be taken earlier, and also choosing $\mu_p=0$ we have
\be
\label{Bdisrel2}
B_{\epi\Lambda}(s,t) = a(t) +  \sum_\ell \int \d \mu \[ \frac{s^2}{\mu-s}+\frac{(-s-t)^2}{\mu+s+t}   \] \frac{\mu \ri_{\ell,\ai}(\mu)}{C_\ell^{(\ai)}(1)} C^{(\ai)}_{\ell}\(1+\frac{2t}{\mu}\)    ,
\ee

To see the simplest examples of these positivity bounds, we may define
\be
f^{(2N,M)}\equiv \frac{1}{2(2N+2)!} \pd_t^{M} \pd_s^{2N+2}B_{\epi\Lambda}(s,t) |_{s,t\to0} .
\ee
Making use of dispersion relation (\ref{Bdisrel2}), we have
\bal
f^{(2N,0)} &= \sum_\ell \int \d \mu \ri_{\ell,\ai}(\mu)   \frac{1}{\mu^{2N}}  >0 ,~~~~N=0,1,2,...   ,
\eal
which are positive, and $f^{(2N-1,0)}=0$ for $N=1,2,3,...$. Making connection to the triple symmetric expansion coefficient $a_{i,j}$ defined in \eref{adef}, we have $f^{(2N,0)} =  {a_{N+1,0}}/{2}$ and so
\be
a_{N,0}>0 ~~{\rm for}~~ N=1,2,... \  .
\ee
Now, we can define an ``expected value'' or ``moment'' over the ``distribution'' $\ri_{\ell,\ai}(\mu)$:\footnote{The significance of the moment of the positive distribution has been emphasized by \cite{AHH}.}
\be
\<\!\<    X(\mu,\ell)  \>\!\> = \frac{\sum_\ell \int \d \mu {\ri_{\ell,\ai}(\mu)}  X(\mu,\ell)}{\sum_\ell \int \d \mu {\ri_{\ell,\ai}(\mu)}  }    .
\ee
We will see that, since the scattering amplitude can be directly linked to this expected value, inequalities associated with generic expected values can be used to derive positivity bounds on the amplitude.

\subsection{Nonlinear positivity bounds with $s$ derivatives only}
\label{sec:nonlns}

 We first look for new positivity bounds with only $s$ derivatives on the amplitude. For this case, we consider $X(\mu,l)={1}/{\mu^{2N}}$ and we have $\<\!\< {1}/{\mu^{2N}}\>\!\>  =  {f^{(2N,0)}}/{f^{(0,0)}}$. Then the Cauchy-Schwarz inequality for expected values,
\be
\dla\frac{1}{\mu^{2I}}\dra \dla\frac{1}{\mu^{2J}}\dra \geq  \dla\frac{1}{\mu^{I+J}}\dra^2    ,
\ee
leads to
\be
f^{(2I,0)}f^{(2J,0)}\geq (f^{(I+J,0)})^2  ,
\ee
or, in terms of the coefficients $a_{N+1,0}$,
\be
\label{nonaIJ}
a_{2I,0}a_{2J,0}\geq (a_{I+J,0})^2    .
\ee
Note that when $I+J$ is an odd number, $f^{(I+J,0)}=0$, so non-trivial constraints come from when $I+J$ is even. In particular, we have
\bal
f^{(4N,0)}f^{(0,0)} \geq (f^{(2N,0)})^2    .
\eal
These positivity constraints are different from the positivity bounds defined in \cite{deRham:2017avq} in that these positivity bounds are nonlinear in the amplitude and its $s$ and $t$ derivatives, while the $Y$ positivity bounds are linear in them.  We would like to mention that the nonlinear positivity bound (\ref{nonaIJ}) has been found previously \cite{AHH}, which was derived in the case of a weakly coupled UV completion by matching to the UV heavy masses and realizing that the $a_{i,0}$ coefficients must be constrained by the convex hull of the half moment curve of the heavy masses. Moreover, \cite{AHH} has shown that the determinant of the Hankel matrix formed by $a_{i,0}$ to a given order is positive. The positivity of this Hankel matrix has found interesting applications in the weak gravity conjecture \cite{Chen:2019qvr} and string theory amplitudes \cite{Green:2019tpt, Huang:2020nqy}. The optimal positivity of the $s$ derivative amplitudes has been recently sought after in \cite{Bellazzini:2020cot} with arc dispersion relations and positive moments, confirming the results of \cite{AHH} and obtaining optimal bounds truncated to a given order.

There are also some other immediate inequalities for expected values. Viewing $1/\mu$ as the independent variable, $(1/\mu)^{N}$ for $N>1$ is a convex function, and so we have Jensen's inequality
\be
\dla \f{1}{\mu^{2N}} \dra \geq \dla \f{1}{\mu^{2}} \dra^{N}    ,
\ee
which leads to
\be
\label{jensenN}
f^{(2N,0)}  \geq \frac{(f^{(2,0)})^N}{(f^{(0,0)})^{N-1}}   .
\ee
This can also be obtained by repeated use of the Cauchy-Schwarz inequality. In general, we have
\be \label{bound100}
f^{(2N,0)}\geq \frac{(f^{(2I,0)})^{J}}{(f^{(0,0)})^{J-1}},~~~N=IJ, ~~N,I=0,1,2,...  ~ .
\ee
There is also Holder's equality, which in terms of $f^{(N,0)}$ is given by
\be
  \(\frac{f^{(2I,0)}}{f^{(0,0)}}\)^{\frac1i} \(\frac{f^{(2J,0)}}{f^{(0,0)}}\)^\frac1j \geq  \frac{f^{(\frac{2I}{i}+\frac{2J}{j},0)}}{f^{(0,0)}} ,~~~~
  \frac1i + \frac1j = 1,~~i,j>1   .
\ee
The Cauchy-Schwarz inequality is the special case where $i=j=2$.

We can already learn something very powerful from for example the simple statement \eqref{jensenN}. If it is ever the case that $f^{(0,0)} / f^{(2,0)} \ll (\epi\Lambda)^{4}$ for example, then it clearly follows that $f^{(2N,0)}/f^{(0,0)}$ becomes arbitrarily large at large $N$, undermining the typical expectations of a low energy EFT expansion. This argument alone does not rule out EFTs for which this is true, but clearly highlights a significant issue. Indeed this argument may be extended for any pair of $f^{(2l,0)}$ as for example from \eref{bound100}. This comes close to ruling out situations where there is a soft behaviour in the amplitude \cite{Cheung:2014dqa,Cheung:2016drk,Bellazzini:2016xrt,Hinterbichler:2014cwa} from admitting a standard UV completion. However, it does not quite achieve this as it only applies at present at $t=0$ which excludes the most interesting case of the Galileon. This will be dealt with by a more refined argument in the next sections.

\subsection{Triple crossing and $t$ derivatives}
\label{sec:tricrotder}

To extract new positivity bounds with $t$ derivatives, we can make use of detailed properties of the Gegenbauer polynomial and the fact that a scalar amplitude is trivially triple crossing symmetric. The dispersion relation (\ref{orignalBrel}) is manifestly $s\leftrightarrow u$ crossing symmetric $B(s,t)=B(u,t)$. Triple crossing symmetry means that $B(s,t)$ should also be $s\leftrightarrow t$ crossing symmetric $B(s,t)=B(t,s)$, which one can impose as a condition on \eref{orignalBrel}. Being more precise, in the case where there scattering states are massive and their is a mass gap to the branch cut, the scattering amplitude will be an analytic function in the so-called Mandelstam triangle, for which the $s$ and $t$ channel dispersion relations may be identified
\begin{eqnarray}
&& a(t) +  \int^{\infty}_{4m^2} \frac{\d \mu}{\pi (\mu-\mu_p)^2} \[ \frac{(s-\mu_p)^2}{\mu-s}+\frac{(u-\mu_p)^2}{\mu-u}   \]   \im A(\mu,t) \nonumber \\
&&  \smallskip=a(s) +  \int^{\infty}_{4m^2} \frac{\d \mu}{\pi (\mu-\mu_p)^2} \[ \frac{(t-\mu_p)^2}{\mu-t}+\frac{(u-\mu_p)^2}{\mu-u}   \]   \im A(\mu,s) \, .
\end{eqnarray}
Note that this relation is not valid outside of the Mandelstam triangle in general.

The $\epi\Lambda$ subtracted amplitude (\ref{Bdisrel}) that is used in the improved positivity bounds is in general not triple crossing symmetric, because the $4m^2$ to $(\epi\Lambda)^2$ subtraction is only $s\leftrightarrow u$ crossing symmetric. Nevertheless, when there is a weakly coupled tree level UV completion, the dispersion relation for the tree level amplitude $B_{\rm tr}(s,t)$ is triple crossing symmetric, as the $4m^2$ to $(\epi\Lambda)^2$ subtraction vanishes then. With this in mind, triple crossing becomes most powerful in the case of weakly coupled tree level UV completions.

To proceed and to simplify the core argument we shall assume $m\ll \Lambda$ and neglect the mass dependence in the partial wave formula, as appropriate for weakly coupled UV completions for which the leading bounds are on the tree amplitudes. Imposing $s\leftrightarrow t$ crossing symmetry at $s=0$, that is, $B_{\rm tr}(0,t) = B_{\rm tr}(t,0)$, we can express the unknown subtraction function $a(t)$ in terms of the dispersion integral:
\bal
 a(t) &=
 a(0) +  \sum_\ell \int \d \mu \Bigg( \[ \frac{t^2}{\mu-t}+\frac{t^2}{\mu+t}   \]   \mu\ri_{\ell,\ai}(\mu)
  - \frac{t^2}{\mu+t}   \frac{\mu\ri_{\ell,\ai}(\mu)}{C^{(\ai)}_{\ell}\(1\)} C^{(\ai)}_{\ell}\(1+\frac{2t}{\mu}\) \Bigg)   .
\eal
Imposing the $s \leftrightarrow t$ crossing symmetry in general and then expanding in terms of powers of kinematic invariants (which amounts to an expansion in $1/\mu$) gives rise to
\bal
\label{stRelation1}
0&=B_{\rm tr}(t,s) - B_{\rm tr}(s,t) = \sum_\ell\int\d \mu \; \ri_{\ell,\ai}(\mu)  \[
\frac{2H_{D,\ell}  st(s^2-t^2)}{(D-2) D   \mu ^2}
+\mc{O}\left(\f1{\mu^3} \right)  \]  ,
\eal
where we have defined
\bal
H_{D,\ell} &= \ell(\ell+D-3) [ 4 - 5 D - 2(3 -  D) \ell + 2 \ell^2]     .
\eal
Since this relation must be true for any $s$ and $t$, it follows that
\bal
\label{crossiden1D}
\sum_\ell\int\d \mu \; \ri_{\ell,\ai}(\mu)   \frac{H_{D,\ell}}{\mu^2}    &=0    .
\eal
must hold as an identity. This is one of the many nontrivial consequences of full crossing symmetry on the partial wave expansion coefficients, which will be explored systematically in Section \ref{sec:triple}. For now, as we shall see, the condition \eref{crossiden1D} already turns out to be remarkably fruitful.

Using the $s\leftrightarrow u$ symmetric dispersion relation, we can cast the amplitude in a triple-crossing-symmetric way $B_{\rm tr}(s,t)= (B_{\rm tr}(s,t)+B_{\rm tr}(s,u)+B_{\rm tr}(t,s))/3$. A straightforward evaluation gives
\iffalse%%%%%%%%%%
\bal
\pd_s^2 C & = 4 \sum_\ell \int \d \mu  {\ri_{\ell,\ai}(\mu)}
\\
\pd_t\pd_s^2 C &= \frac{2}{D-2}\sum_\ell \int \d \mu  {\ri_{\ell,\ai}(\mu)}    [ 3(2 -  D) + 4(-3 + D)\ell + 4 \ell^2 ]
\\
\pd_t^2\pd_s^2 C &= 24 \sum_\ell \int \d \mu \frac{\ri_{\ell,\ai}(\mu)}{\mu^2}
\\
\pd_s^4 C &= 48 \sum_\ell \int \d \mu \frac{\ri_{\ell,\ai}(\mu)}{\mu^2}
\eal
\fi%%%%%%%%%%%%
\bal
\label{f01rel}
\frac{f^{(0,1)}}{f^{(0,0)}} &=   \dla  \frac{3(2 -  D) + 4(-3 + D) \ell + 4 \ell^2}{2(D-2)\mu}  \dra    ,
\eal
which leads to
\be
\frac{f^{(0,1)}}{f^{(0,0)}}+  \dla \f{3}{2\mu} \dra  =      \dla  \frac{ 2 (-3 + D) \ell + 2 \ell^2}{(D-2)\mu}  \dra  .
\ee
A special case of the Cauchy-Schwarz inequality of the expected values  $\<\!\< X(\mu,l)\>\!\>^2 \leq  \<\!\< X(\mu,l)^2 \>\!\>$ (or ``the variance is positive'') tells us that
\be
\label{varispos}
\(\frac{f^{(0,1)}}{f^{(0,0)}}+   \dla \f{3}{2\mu} \dra \)^2  = \dla  \frac{ 2( D-3) \ell + 2 \ell^2}{(D-2)\mu} \dra^2 \leq  \dla \( \frac{ 2( D-3) \ell + 2 \ell^2}{(D-2)\mu}  \)^2 \dra   .
\ee
Since we can split the square into
\be
 ( 2( D-3) \ell + 2 \ell^2)^2 =( 5 D-4)  \[2( D-3) \ell + 2 \ell^2 \]  +2 H_{D,\ell}    ,
\ee
plugging back into \ref{varispos}, the later term vanishes due to Eq \ref{crossiden1D},  so we get
\bal
\label{ineqC}
\(\frac{f^{(0,1)}}{f^{(0,0)}}+   \dla \f{3}{2\mu} \dra\)^2 & \leq  \frac{5 D-4}{D-2}  \dla \frac{ 2 ( D-3)  \ell + 2  \ell^2}{(D-2)\mu^2}   \dra     .
\eal
Note that the integrand of the integral $\sum_\ell \int \d \mu {\ri_{\ell,\ai}(\mu)}(...) /\mu^2$ is positive definite. So if one fixes one of the $\mu$'s in the denominator to the lower limit of the integration, which is $\Lambda_{\rm th}^2$ for this case,  the result is greater than the original integral. For the case where $\Lambda_{\rm th}=\Lambda$, we have the following inequality
\be
\dla \frac{ 2 ( D-3)  \ell + 2  \ell^2}{(D-2)\mu^2}   \dra    <  \frac{1}{\Lambda^2}  \dla \frac{ 2 ( D-3)  \ell + 2  \ell^2}{(D-2)\mu}   \dra   .
\ee
Combining it with \eref{ineqC}, we have
\be
\(\frac{f^{(0,1)}}{f^{(0,0)}}+   \dla \f{3}{2\mu} \dra\)^2  <  \frac{5 D-4}{(D-2)\Li^2} \(\frac{f^{(0,1)}}{f^{(0,0)}} +   \dla \f{3}{2\mu} \dra  \)   ,
\ee
which can be written as
\be
\label{tbound1}
0< \f{f^{(0,1)}}{f^{(0,0)}} +   \dla \f{3}{2\mu} \dra  <    \frac{5 D-4}{(D-2)\Li^2}   .
\ee
Since $ \<\!\< {1}/{\mu} \>\!\>$ and $f^{(0,0)}$ are positive, we have
\be
\label{tbound2}
 f^{(0,1)}  <    \frac{5 D-4}{(D-2)\Li^2} f^{(0,0)}
\ee
Similarly, we have the inequality
\be
 \dla \f{1}{\mu} \dra   <  \f1{\Lambda^2}  \marrow   \sum_\ell \int \d \mu \ri_{\ell,\ai}(\mu)   \frac{1}{\mu} <    \f1{\Lambda^2}  f^{(0,0)}   ,
\ee
and thus we have
\be
\label{tbound3}
0< f^{(0,1)}+  \f3{2\Lambda^2}   f^{(0,0)} .
\ee
In other words we have both an upper and lower bound on $f^{(0,1)}$, bounded by a term of the same order.
\be \label{upperlower}
- \f3{2\Lambda^2}   f^{(0,0)} <f^{(0,1)}<\frac{5 D-4}{(D-2)\Li^2} f^{(0,0)}
\ee
This is a remarkably strong restriction on the parameter of the effective theory.

We will generalize these new positivity bounds in Section \ref{sec:npbG}, following a similar argument, and compare the new positivity bounds to the previous $Y$ bounds in Section \ref{sec:comp}, but before that, as an example of potential applications, we will show that these first new bounds indeed provide extra constraints on an EFT and already have important implication for weakly broken Galileons in the next section.

\section{Implication for weakly broken Galileon theories}
\label{sec:galileon}

In this section, we apply the new positivity bounds derived above to weaklly broken Galileon theory. The $Y$ positivity bounds \cite{deRham:2017avq} were applied to the specific case of a massive Galileon in \cite{deRham:2017imi} and it was found that there is a parameter region where the theory is compatible with analyticity. We will see that the new positivity bounds can rule out massive Galileon and more generally any weakly broken Galileon theory as an EFT with a healthy hierarchy and standard local UV completion.

Let us first see what the nonlinear forward limit positivity bounds imply for a generic scattering amplitude which is parametrized as in Eq (\ref{adef}) and where $\tilde{a}_{1,0}$ is suppressed $\tilde{a}_{1,0}\sim g^2$ with $g\ll 1$. With the notation (\ref{adef}), the positivity bound (\ref{jensenN}) can be written as
\be
\label{aNp1}
\tilde{a}_{N+1,0}\, \tilde{a}_{1,0}^{N-1}\geq \tilde{a}_{2,0}^N     .
\ee
In the large $N$ limit, this bound essentially implies $\tilde{a}_{1,0} > \tilde{a}_{2,0}$ if we assume the high order coefficients do not grow arbitrarily large\;\footnote{This assumption can actually be dropped; see \eref{c1c2c3}  with $\epi=1$. Note that $c_{m,0}=a_{m,0}$.}. Since $\tilde{a}_{N,0}> 0$ for $N=1,2,3,...$, the fact that $\tilde{a}_{1,0}$ is suppressed ($\tilde{a}_{1,0}\sim g^2$) then implies that $\tilde{a}_{2,0}$ also has to be suppressed ($\tilde{a}_{2,0}\sim g^2$). With this established, we can go back to \eref{aNp1}, and we can then infer that generically $\tilde{a}_{N,0}\sim g^2$. So in the forward limit $t=0$, neglecting the constant term, the amplitude should schematically go like
\be
B(s,0) \sim  \frac{g^2}{\Lambda^{D-4}} \(  \frac{x}{\Lambda^{4}}     +  \frac{x^2}{\Lambda^{8}}  +  \frac{x^3}{\Lambda^{12}}  + \cdots  \)    ,
\ee
where we have neglected order unity coefficients. This is as far as we get with the forward limit bounds, but does not say anything interesting about the Galileon case for which it is the leading $y \sim - s t u $ term that is relevant.

On the other hand, the $t$ derivative positivity bound (\ref{tbound2}) implies
\be
  \tilde{a}_{0,1}   < \frac{5 D-4}{(D-2)} \tilde{a}_{1,0} ,
\ee
and the bound (\ref{tbound3}) implies
\be
0<  \tilde{a}_{0,1} + \f32  \tilde{a}_{1,0}   .
\ee
Since $\tilde{a}_{1,0}$ is suppressed $\tilde{a}_{1,0}\sim g^2$, the two inequalities above imply that $\tilde{a}_{0,1}$ also has to be suppressed, that is, $\tilde{a}_{0,1}\sim g^2$. Therefore, for an amplitude where the leading term $\tilde{a}_{1,0}$ is soft, positivity bounds implies that the amplitude has to be of the schematic form
\be
\label{Bggen}
B(s,t) \sim   \frac{g^2}{\Lambda^{D-4}} \(  \frac{x}{\Lambda^{4}}   +   \frac{y}{\Lambda^{6}}  +  \frac{x^2}{\Lambda^{8}}    + \cdots  \)   ,
\ee
again neglecting order unity factors.

Galileon theories are scalar field theory that captures salient features of a number of massive gravitational models  \cite{Dvali:2000hr, deRham:2010eu,deRham:2010ik, deRham:2010kj} but have also been considered in their own right as effective theories with soft behaviour for their scattering amplitudes \cite{Nicolis:2008in}. The Galileon symmetry $\pi \rightarrow \pi + b_{\mu} x^{\mu}$ translates directly into the requirement that the usual $\mc{O}(E^4)$ term in the scattering amplitude low energy expansion vanishes. When the Galileon symmetry is weakly broken, the term is recovered but with a coefficient which is suppressed by the amount of breaking. In $D$ dimensions, it is given by the following Lagrangian
\bal
\Lambda_3^{4-D}\mc{L}_{\text{mg}}
&= - \frac{1}{2} \partial_\mu \pi \pd^\mu \pi - \frac12 m^2 \pi^2  + \sum_{n=3}^{D+1}\f{g_n}{\Lambda_3^{3n-3}} \pi \pd^{\mu_1}\pd_{[\mu_1}\pi \pd^{\mu_2}\pd_{\mu_2}\pi\cdots \pd^{\mu_{n}}\pd_{\mu_{n}]}\pi
\nn
&~~~~~ + \sum_i \mc{O}_i\(\f{\pd^2\pi}{\Lambda_3^3},\f{\pd^3\pi}{\Lambda_3^4},\f{\pd^4\pi}{\Lambda_3^5},...\)   ,
\eal
where $g_n$ are dimensionless coefficients of order one, ${}_{[~]}$ is anti-symmetrization of the indices,  $\Lambda_3$ is the strong coupling scale and the $\mc{O}_i$ operators represent higher derivative terms, which if not present at the classical level can be generated by quantum corrections. An explicit calculation shows that the scattering amplitude for massive galileon goes like \cite{deRham:2017imi}
\be \label{Bform100}
B_{\rm mg}(s,t) \sim \frac{1}{\Lambda_3^{D-4}} \(  \f{m^2}{\Lambda_3^6} x   +  \f{1}{\Lambda_3^6} y +  \f{1}{\Lambda_3^{8}}  x^2 + \cdots \)     ,
\ee
where we have neglected the constant term of order ${m^2}/{\Lambda_3^2}$ and order unity coefficients. So the massive galileon amplitude belongs to the type of the amplitude where the leading low energy behaviour is soft. Matching this amplitude to Eq (\ref{Bggen}), we can infer that
\be
\f{g^2}{\Li^D} \sim \f{m^2}{\Lambda_3^{D+2}},~~~\f{g^2}{\Li^{D+2}} \sim \f{1}{\Lambda_3^{D+2}}    .
\ee
Therefore, the new positivity bounds imply that for massive galileon the cutoff of the theory has to be parametrically close to the mass of the field
\be
\Li \sim m  ,
\ee
which contradicts the most basic requirement of an EFT, a healthy hierarchy between the two scales.

A similar argument applies a theory of a massless Galileon with a small Galileon symmetry breaking term such as
\bal
\Lambda_3^{4-D}\mc{L}_{\text{wbg}}
&= - \frac{1}{2} \partial_\mu \pi \pd^\mu \pi  - \frac{\alpha_4}{\Lambda_3^4} (\partial \pi)^4 + \sum_{n=3}^{D+1}\f{g_n}{\Lambda_3^{3n-3}} \pi \pd^{\mu_1}\pd_{[\mu_1}\pi \pd^{\mu_2}\pd_{\mu_2}\pi\cdots \pd^{\mu_{n}}\pd_{\mu_{n}]}\pi
\nn
&~~~~~ + \sum_i \mc{O}_i\(\f{\pd^2\pi}{\Lambda_3^3},\f{\pd^3\pi}{\Lambda_3^4},\f{\pd^4\pi}{\Lambda_3^5},...\)   ,
\eal
with $|\alpha_4| \ll 1$ the measure of the Galileon symmetry breaking. The form of the scattering amplitude is then again \eqref{Bform100} with now $m^2 \rightarrow \alpha_4 \Lambda_3^2$, and the bounds \eqref{tbound2} and \eqref{tbound3} amount to the requirement that $|\alpha_4| \sim {\cal O}(1)$ (see \cite{Bellazzini:2017fep} for a weaker lower bound on $|\ai_4|$).  Stated differently, the Galileon symmetry can never just be weakly broken, since this would require the leading $x$ term to be suppressed relative to the $y$ term in a way forbidden by the new positivity bounds.

\section{New positivity bounds: Generalizations}
\label{sec:npbG}

Having seen how powerful these new positivity bounds can be in the example in the last section, in this section, we will generalize the new positivity bounds found in Section \ref{sec:newBoundExample}, again making use of the finer properties of the partial wave expansion and the triple crossing symmetry that are already exploited there, but in a more systematical way.

Note that dispersion relation \eref{Bdisrel21} is $s\leftrightarrow u$ symmetric, so we may take advantage of an $s\leftrightarrow u$ symmetric variable
\be
w \= -\hat s\hat u=\hat s(\hat s+ t),~~~~\hat s=s-2m^2,~~~~\hat u=u-2m^2
\ee
So, slightly different from the way to derive the $Y^{(2N,M)}$ bounds or the direct approach in Section \ref{sec:newBoundExample}, here we find it convenient to expand the amplitude in terms of $w$ and $t$. Choosing the subtraction point at $\mu_p=2m^2$ and defining $\hat\mu=\mu-2m^2$, the dispersion relation can be re-cast as\footnote{Note that the $\mu$ here differs by a shift of $2m^2$ from the $\mu$ previously.}
\bal
B_{\epi\Lambda}(s,t)&=a(t)+\sum_\ell \int \d\mu \rho_{\ell,\alpha}(\mu+2m^2)\mu
\[ \frac{\hat s^2}{\mu-\hat s}+\frac{\hat u^2}{\mu-\hat u} \]
\frac{C_\ell^{(\alpha)}(1+\frac{2t}{\hat\mu})}{2C_\ell^{(\alpha)}(1)}\\
&={b}(t)+\sum_\ell \int \d\mu \rho_{\ell,\alpha}(\mu+2m^2)\mu^2
\frac{2+\frac{t}{\mu}}{1+\frac{t}{\mu}-\frac{w}{\mu^2}}
\frac{C_\ell^{(\alpha)}(1+\frac{2t}{\hat\mu})}{2C_\ell^{(\alpha)}(1)}  ,
\eal
where now the integration limit of $\mu$ is from $(\epsilon\Lambda)^2-2m^2$ to infinity and $b(t)$, although not important for this paper, is for clarity given by
\be
b(t)=a(t) +\sum_\ell \int \d\mu \rho_{\ell,\alpha}(\mu+2m^2)
[t\mu-2\mu^2]
\frac{C_\ell^{(\alpha)}(1+\frac{2t}{\hat\mu})}{2C_\ell^{(\alpha)}(1)} .
\ee
Expanding $w/\mu^2$ in the denominator and $2t/\hat\mu$ in $C_\ell^{(\alpha)}(1+\frac{2t}{\hat\mu})$ respectively and then expanding $t/\mu$ in the the denominator, we can get
\beq
\label{bcwt}
B_{\epi\Lambda}(s,t)= \sum_{m=0}^{\infty} \sum_{n=0}^{\infty} c_{m,n} w^m t^n  ,
\eeq
where the expansion coefficients are given by
\bal
\label{cml}
c_{m,n}
& \= \<\frac{\hat D_{m,n}}{\mu^{2m+n-2}}\> \= \sum_{i=0}^{n} (-1)^i \mtM^i_m
\< \frac{\mtL^{n-i}_\ell}{\mu^{2m+n-2}}\>\\
&=\< \frac{\mtL^{n}_\ell}{\mu^{2m+n-2}}\>-\mtM^1_m \<\frac{\mtL^{n-1}_\ell}{\mu^{2m+n-2}}\>
+\mtM^2_m \< \frac{\mtL^{n-2}_\ell}{\mu^{2m+n-2}}\>+...\ .
\eal
and for $c_{0,n}$ we also need to add the Taylor coefficients of $b(t)$, although the $c_{0,n}$ terms will play no essential role in the following. Here $\mtL^n_\ell$, up to a factor, are semi-positive polynomials of $\ell$, defined as
\bal
\label{linko}
\mtL^n_\ell  \= \tilde\mtL^n_\ell \(\f{\mu}{\hat\mu}\)^n, ~~~~ \tilde\mtL^n_\ell \= \frac{1}{n!}\frac{\Gamma(\ell+n+2\alpha)}{\Gamma(\ell+2\alpha)}
\frac{\Gamma(\ell+1)}{\Gamma(\ell-n+1)}\frac{\Gamma(\alpha+\frac{1}{2})}{\Gamma(\alpha+\frac{1}{2}+n)}
\geq 0  ,
\eal
while $\mtM^i_m$ are positive numbers defined as
\bal
\label{mareo}
\mtM^i_m \= \frac{1}{2}\left[\frac{(m-1+i)!}{(m-1)!i!}+\frac{(m+i)!}{m!i!}\right] > 0  .
\eal
The average $\<~\>$ is defined as
\be
\label{averagedef}
\< X(\mu,\ell)\> = \sum_\ell \int \d \mu \rho_{\ell,\alpha}(\mu+2m^2) X(\mu,\ell) ,
\ee
which is different from the normalized average $\<\!\<~\>\!\>$ in Section \ref{sec:newBoundExample}. With this average, $c_{m,0}$ can be written as
\beq
c_{m,0}=\< \frac{1}{\mu^{2(m-1)}}\>   .
\eeq
By limiting factors of $\mu$ in the denominator to be the lower limit of the integration $(\epi\Lambda)^2$, we immediately get that
\be
\label{c1c2c3}
c_{1,0}>(\epi\Lambda)^2 c_{2,0}>(\epi\Lambda)^4 c_{3,0}>(\epi\Lambda)^6 c_{4,0}>...\, .
\ee
To translate between the $s\leftrightarrow u$ symmetric $c_{m,n}$ and the triple symmetric $a_{i,j}$ parameters, using $x=w+t^2$ and $y=wt$, we can get the relation
\be
\label{cmntoaij}
c_{m,n}=\sum _{k\in \mathbb{N},-3 k+m+n\geq 0,n-2 k\geq 0} \frac{(3 k+m-n)! a_{3 k+m-n,n-2 k}}{k! (2 k+m-n)!} .
\ee
Then the Cauchy-Schwarz nonlinear positivity bounds with only $s$ derivatives in Section \ref{sec:nonlns} can be written as
\beq
\label{ccgcc}
c_{m,0}~c_{m+2,0} > (c_{m+1,0})^2   .
\eeq
As mentioned above, we can also get additional inequalities from \eref{ccgcc} (and $c_{m,0}>0$),  such as $c_{m,0}~c_{n,0} \geq (c_{\frac{m+n}{2},0})^2 ~~ {\rm for} ~ m+n={\rm even}$, and $(c_{m,0})^{n-1}~c_{m+n,0} > (c_{m+1,0})^n$, but the bounds from \eref{ccgcc} are independent and complete, in the sense that those extra bounds can be derived from them.  For example, multiplying $c_{m,0} c_{m+2,0}>c_{m+1,0}^2$ and $c_{m+2,0} c_{m+4,0}>c_{m+3,0}^2$ and the square of $c_{m+1,0} c_{m+3,0}>c_{m+2,0}^2$, we can derive $c_{m,0} c_{m+4,0}>c_{m+2,0}^2$; similarly, we can derive $c_{m,0} c_{m+6,0}>c_{m+3,0}^2$ and so on.  Also, noticing that $\<1/\mu\>^2 < c_{1,0}c_{2,0}$, the $t$-derivative bounds \eref{tbound2} and \eref{tbound3} become
\bal
\label{t1boounc}
-\frac{3}{2}\sqrt{c_{1,0}c_{2,0}} <  c_{1,1} < \frac{10\ai+11}{2\ai+1} \sqrt{c_{1,0}c_{2,0}}   .
\eal

\subsection{The $PQ$ positivity bounds}
\label{sec:nonlinear}

Now, we generalize the $s$-derivative nonlinear positivity bounds to allow for $t$ derivatives. Similar to \cite{deRham:2017avq}, we will do linear cancelation among $c_{m,n}$. The problem, as observed in \cite{deRham:2017avq}, is that we cannot do it at the level of the same $\mu$ power. To overcome this, we have to fix one of the $\mu$ denominator to the lower limit of the integration, which is now
\be
(\epi\hat\Lambda)^2=(\epi\Lambda)^2-2m^2,
\ee
to obtain {\it the relaxing inequality}. If we consider $(\epi\Lambda)^2\gg m^2$, then $(\epi\hat\Lambda)^2\simeq (\epi\Lambda)^2$. Here, as oppose to using it in one direction \cite{deRham:2017avq}, now we use the relaxing inequality in both directions, that is, we will use something like\;\footnote{Note that $L^i_\ell$ can vanish, but that is if and only if $\ell=0$. Since an amplitude must also have at least some of $\ell>0$ waves, we have ``$>$'' instead of ``$\geq$'' in these inequalities.}
\beq
\label{relax0}
\frac{1}{(\epi\hat\Lambda)^2} \< \frac{L^i_\ell}{\mu^{j-1}} \>  > \< \frac{L^i_\ell}{\mu^{j}} \> >
(\epi\hat\Lambda)^2  \< \frac{L^i_\ell}{\mu^{j+1}} \>  .
\eeq
Similar to the case of the $Y$ bounds \cite{deRham:2017avq}, the alternating sign in \eref{cml} is the main obstacle to get positivity for $t$ derivatives. Again, we can perform linear cancellations to overcome this, but now we will introducing two sequences of linear combinations of $c_{m,n}$: $P_{m,n}$ and $Q_{m,n}$.

\subsubsection{The nonlinear $PQ$ bounds}

First, we start with the case with the 1st $t$ derivative. Considering
\be
c_{m,1}=\< \frac{\mtL^1_\ell}{\mu^{2m-1}} \>-\mtM^1_m \< \frac{1}{\mu^{2m-1}} \> ,
\ee
we can define
\beq
\mtP_{m,1} \= c_{m,1}+\frac{\mtM^1_m}{(\epi\hat\Lambda)^2} c_{m,0}  ,
\eeq
\beq
\mtQ_{m,1} \= c_{m,1}+\mtM^1_m(\epi\hat\Lambda)^2 c_{m+1,0}    .
\eeq
Using the relaxing inequality \eref{relax0}, we can obtain
\be
\mtP_{m,1} > \< \frac{\mtL^1_\ell}{\mu^{2m-1}} \> > \mtQ_{m,1}   ,
\ee
We can define a modified (positive) density $\tilde{\rho}^i _{\ell,\alpha}=\rho _{\ell,\alpha}\mtL^i_\ell$, and the Cauchy-Schwarz inequality implies that
\beq
\mtP_{m,1}~\mtP_{m+2,1} >
\< \frac{\mtL^1_\ell}{\mu^{2m-1}} \> \< \frac{\mtL^1_\ell}{\mu^{2m+3}} \>
>
\< \frac{\mtL^1_\ell}{\mu^{2m+1}} \>^2
> \mtQ_{m+1,1}^2   .
\eeq
Similarly, for the 2nd $t$ derivative, we have
\be
c_{m,2} =\left\langle\frac{\mathrm{L}_{\ell}^{2}}{\mu^{2 m}}\right\rangle -{M}_{m}^{1}\left\langle\frac{\mathrm{L}_{\ell}^{1}}{\mu^{2 m}}\right\rangle  +{M}_{m}^{2}\left\langle\frac{1}{\mu^{2 m}}\right\rangle  .
\ee
Making use of the relaxing inequality \eref{relax0} and the lower $t$ derivative bounds $-\<{\mtL^1_\ell}/{\mu^{2m-1}}\> > - P_{m,1}$, $-\<{\mtL^1_\ell}/{\mu^{2m+1}}\> < -Q_{m+1,1}$, we can use a linear combination of $c_{i,0}$ and $c_{i,1}$ to  cancel all the other terms but the leading $\<{\mtL^2_\ell}/{\mu^{2m}}\>$ in $c_{m,2}$. Defining
\bal
P_{m,2}&\equiv c_{m,2}+ \f{\mtM^1_m}{(\epi\hat\Lambda)^2} P_{m,1}-\mtM^2_m c_{m+1,0}   ,% \geq \<\frac{\mtL^2_\ell}{\mu^{2m}}\>
\\
Q_{m,2}&\equiv c_{m,2}+\mtM^1_m(\epi\hat\Lambda)^2 Q_{m+1,1}-\mtM^2_m c_{m+1,0}  , %\leq \<\frac{\mtL^2_\ell}{\mu^{2m}}\>
\eal
we can get
\be
P_{m,2}  > \<\frac{\mtL^2_\ell}{\mu^{2m}}\>  >   Q_{m,2}  ,
\ee
which leads to
\be
\mtP_{m,2}~\mtP_{m+2,2} >  (\mtQ_{m+1,2})^2  .
\ee
For the 3rd $t$ derivative, we have
\be
c_{m,3}=\<\frac{\mtL^3_\ell}{\mu^{2m+1}}\>-\mtM^1_m\<\frac{\mtL^2_\ell}{\mu^{2m+1}}\>
+\mtM^2_m\<\frac{\mtL^1_\ell}{\mu^{2m+1}}\>-\mtM^3_m\<\frac{1}{\mu^{2m+1}}\>  ,
\ee
which inspires us to define
\bal
P_{m,3}&\equiv
c_{m,3}+\frac{\mtM^1_m}{(\epi\hat\Lambda)^2}P_{m,2}-\mtM^2_mQ_{m+1,1}+
\frac{\mtM^3_m}{(\epi\hat\Lambda)^2}c_{m+1,0}  , %\geq \<\frac{\mtL^3_\ell}{\mu^{2m+1}}\>
\\
Q_{m,3}&\equiv c_{m,3}+\mtM^1_m(\epi\hat\Lambda)^2 Q_{m+1,2}-\mtM^2_mP_{m+1,1}+
\mtM^3_m(\epi\hat\Lambda)^2 c_{m+2,0}    , % \leq \<\frac{\mtL^3_\ell}{\mu^{2m+1}}\>
\eal
and we can again get
\be
 P_{m,3}  > \<\frac{\mtL^3_\ell}{\mu^{2m+1}}\> > Q_{m,3}  ,
\ee
which leads to
\be
\mtP_{m,3}~\mtP_{m+2,3} > (\mtQ_{m+1,3})^2 .
\ee
Thus, for generic $t$ derivatives, we need to mix $P_{m,n}$ and $Q_{m,n}$, and considering the structure of \eref{cml}, it is not difficult to find the linear combinations for generic $m$ and $n$:
\bal
\mtP_{m,n}& \= c_{m,n}+\frac{1}{(\epi\hat\Lambda)^2}\sum_{i=1}^{\lfloor\frac{n+1}{2}\rfloor}\mtM^{2i-1}_m
\mtP_{m+i-1,n+1-2i}-\sum_{j=1}^{\lfloor\frac{n}{2}\rfloor} \mtM^{2j}_m\mtQ_{m+j,n-2j}   ,
%\geq \<\frac{\mtL^{n}_\ell}{\mu^{2m+n-2}}\>
\\
\mtQ_{m,n}& \= c_{m,n}+(\epi\hat\Lambda)^2\sum_{i=1}^{\lfloor\frac{n+1}{2}\rfloor} \mtM^{2i-1}_m
\mtQ_{m+i,n+1-2i}-\sum_{j=1}^{\lfloor\frac{n}{2}\rfloor} \mtM^{2j}_m\mtP_{m+j,n-2j}  ,
%\leq \<\frac{\mtL^{n}_\ell}{\mu^{2m+n-2}}\>
\eal
\be
\label{PQinequ}
\mtP_{m,n} > \<\frac{\mtL^{n}_\ell}{\mu^{2m+n-2}}\> > \mtQ_{m,n}   ,
\ee
where we have defined $P_{m,0} = Q_{m,0} = c_{m,0}$ and $\lfloor\;\rfloor$ again means taking the flooring integer. Thus, the simplest of generic nonlinear positivity bounds take the form
\beq
\mtP_{m,n}~\mtP_{m+2,n} > (\mtQ_{m+1,n})^2   .
\eeq
The $n=0$ bounds reduce to the $s$-derivative nonlinear positivity bounds.

\subsubsection{The linear $PQ$ bounds}

Having defined the $\mtP_{m,n}$ and $\mtQ_{m,n}$ quantities, by \eref{PQinequ} and the relaxing inequality,  it is easy to see that they can also be used to construct a set of linear positivity bounds:
\be
\mtP_{m,n} > (\epi\hat\Lambda)^{4k} \mtQ_{m+k,n}, ~~~k=0,1,2,...  ,
\ee
which we will refer to as the linear $PQ$ positivity bounds. Also by \eref{PQinequ}, we can find that
\be
\mtP_{m,n} > 0  ,
\ee
which are very similar to the $Y$ positivity bounds in \cite{deRham:2017avq}. In Section \ref{sec:comp}, we compare the $P>0$ bounds with the $Y$ bounds by explicitly evaluating the first few bounds in the two sets; see Fig \ref{fig:YP}.

\subsection{The $D^{\rm su}$ positivity bounds}
\label{sec:evenmoreb}

In Section \ref{sec:nonlinear}, we have used the relaxing inequality and the Cauchy-Schwarz inequality
to obtain the linear and nonlinear $PQ$ positivity bounds. Here and in Section \ref{sec:triple} we make use of a new strategy to extract more positivity bounds from the partial wave expanded dispersion relation. To understand the basic idea, we first note that since $1/\mu<1/\hat\mu<\si/\mu$ with
\be
\label{sidef}
\si=\f{(\epi\hat\Lambda)^2}{(\epi\hat\Lambda)^2-2m^2},
\ee
we have
\be
\label{ctoD}
c_{m,n}= \<\f{\hat D_{m,n}}{\mu^{2m+n-2}}\> >   \< \f{ D_{m,n}}{\mu^{2m+n-2}} \>   .
\ee
where we have defined
\be
\label{Dk+}
D_{m, n} \= \sum_{{\rm even} ~i= 0}^n M_{m}^{i} \tilde L_{\ell}^{n-i} -\sum_{{\rm odd} ~i= 1}^n  \sigma^{n-i} M_{m}^{i} \tilde L_{\ell}^{n-i}  .
\ee
When $(\epi\Lambda)^2\gg m^2$, $\hat D_{m, n}$ simply becomes $D_{m, n}$. Since $\<~\>$ is an integration over a positive density distribution $\rho _{\ell,\alpha}$, bounds on $c_{m,n}$ can be extracted if $D_{m,n}$ is bounded below for all possible $\ell$. Indeed, $D_{m,n}$ is bounded below as it is a $2n$-th order polynomial in term of $\ell$ with a positive coefficient for the highest order.

To see this, note that $\tilde\mtL^n_\ell$ are roughly speaking the derivatives of the Gegenbauer polynomial $C^{(\ai)}_{\ell}(x)$ evaluated at $x=1$, which are $n$-th order polynomials in terms of $\ell$:
\bal
\tilde\mtL^n_\ell &=\frac{1}{n!}\frac{\Gamma(\ai+\frac{1}{2})}{\Gamma(\ai+\frac{1}{2}+n)}
\ell(\ell+2\ai)~(\ell-1)(\ell+2\ai+1)... (\ell-n+1)(\ell+2\ai+n-1)  .
\eal
Inspecting the structure of $\tilde\mtL^n_\ell$, it is convenient to introduce a new variable
\be
\label{etadef}
\ei \= \ell(\ell+2\ai)
\ee
and then $\tilde\mtL^n_\ell$ can be cast as $n$-th order polynomials in terms of $\ei$:
\bal
\tilde\mtL^n_\ell &=\frac{1}{n!}\frac{\Gamma(\ai+\frac{1}{2})}{\Gamma(\ai+\frac{1}{2}+n)}
\prod_{k=0}^{n-1}(\ei-k(2\ai+k))  .
\eal
Note that $\ei$ is always non-negative $\ei\geq 0$ for $\ai\geq 0$, ie, for $D\geq 3$. Then we can re-cast $D_{m, n}$ as
\be
\label{bibeo}
D_{m, n}(\eta) \= \sum_{i=0}^{n}(-1)^i \mtW^i_{m,n} \ei^{n-i}
\ee
where
\beq
\mtW^i_{m,n}\equiv  \sum_{{\rm even}~k=0}^{i}\mtM^{k}_m  R_{n-k}  \ai^{n-k-1}_k + \sum_{{\rm odd}~k=1}^{i} \si^{n-k}\mtM^{k}_m  R_{n-k}  \ai^{n-k-1}_k  > 0   ,
\eeq
with
\bal
R_{n} &\= \frac{1}{n!}\frac{\Gamma(\ai+\frac{1}{2})}{\Gamma(\ai+\frac{1}{2}+n)} ,
\\
 \ai^l_k &\= \sum_{0 \leq i_1 \neq...\neq i_k \leq l}
i_1(2\ai+i_1)...i_k(2\ai+i_k),~~~~\ai^l_0\equiv 1  .
\eal
From \eref{bibeo}, we see that $D_{m,n}(\ei)$ is an $n$-th order polynomial with a positive coefficient for the highest order term, so $D_{m,n}(\ei)$ must be bounded below for non-negative $\eta$, and thus by replacing $D_{m,n}(\ei)$ with its minimum in \eref{ctoD}, we can get a lower bound for $c_{m,n}$.
Suppose $S^0_{m,n}$ is the minimum of $D_{m,n}(\ei)$ over all possible values of $\eta$
\be
S^0_{m,n}= {\rm min}_\ei\; D_{m,n}(\ei) .
\ee
By \eref{ctoD}, we immediately get the lower bounds for $n=2k$
\be
\label{lobod0}
c_{m,2k} - S^0_{m,2k}~c_{m+k,0} > 0   .
\ee
For the case of an odd number of $t$ derivatives (that is, $n=2k+1$), the $\eta^0$ term in $D_{m,n}(\ei)$ is negative, so $S^0_{m,n}$ must be negative, which leads to
\bal
\label{oddbound52}
c_{m,2k+1}  -   S^0_{m,2k+1} \sqrt{c_{m+k,0}c_{m+k+1,0}} >  c_{m,2k+1}-S^0_{m,2k+1} \<\frac{1}{\mu^{2m+2k-1}}\> > 0   .
\eal
From \eref{etadef}, we see that $\ei$ can choose only discrete values. Nevertheless, we may also extend $\eta$ to the real domain and simply find the minimum of $D_{m,n}(\ei)$ over $\eta>0$, denoted as $S^{0\rm R}_{m,n}=D_{m,n}(\ei_R)$. If one directly takes $S^{0\rm R}_{m,n}$ in \eref{lobod} or \eref{oddbound52}, one still gets a valid positivity bound, albeit slightly weaker than using $S^0_{m,n}$ by choosing the allowed discrete $\eta$.

It would be instructive to look at a simple example: the case of $D_{2,2}(\eta)$ in the limit of $\si=1$. Evaluating \eref{bibeo} directly, we can get
\be
D_{2,2}(\eta)=\frac{1}{2 \(\ai+\frac{1}{2}\) \(\ai+\frac{3}{2}\)}
\( \ei^2-\frac{14\ai+17}{2}\ei+\frac{36\ai^2+72\ai+27}{4} \)  .
\ee
The minimum of $D_{2,2}(\eta)$ in the positive real domain is
\be
S^{0\rm R}_{2,2}  = -\frac{52\ai^2+188\ai+181}{32\(\ai+\frac{1}{2}\) \(\ai+\frac{3}{2}\)} \leq  D_{2,2}(\eta)  ,
\ee
which is the value of $D_{2,2}(\ell(\ell+2\ai))$ at $\ell_R=\frac{1}{2} \sqrt{4 \alpha ^2+14 \alpha +17}-2 \alpha$. So we have the positivity bound
\be
c_{2,2}+\frac{52\ai^2+188\ai+181}{32\(\ai+\frac{1}{2}\)\(\ai+\frac{3}{2}\)}c_{3,0} > 0   .
\ee
In 4D ($\ai=1/2$), $\ell_R=2$ is an integer, so we have $S^0_{2,2}=S^{0\rm R}_{2,2}$ and the positivity bound is simply $c_{2,2}+\f92 c_{3,0}\geq 0$. In 5D, however, $\ell_R\simeq 1.96$, and a better bound can be obtained by using $S^0_{m,n}$. To this end, we compute $D_{2,2}(\ell(\ell+2))$ for all the even $\ell$ and find that $S^0_{m,n}=D_{2,2}(2(2+1))=-12/5$, which leads to a slightly stronger bound $c_{2,2}+\f{12}{5} c_{3,0} > 0$.

There is an obvious generalization to these positivity bounds. Instead of extracting the positivity bounds from \eref{ctoD} with the minimum of $D_{m,n}$, we may also linearly superimpose $c_{m,n}$ at the same level of $1/\mu$. That is, we consider a linear combination of the form
\be
\label{mcDbound}
{D}^{\rm su}_{m,n} (\eta,k) \equiv   \sum_{i\geq 0} k_i D_{m+ i,n-2i}(\eta) , ~~~k_0=1 .
\ee
which is still an $n$-th order polynomial of $\eta$ with a positive coefficient for the highest order term. (For example, for $D_{3,3}(\eta)$, we can consider ${D}^{\rm su}_{3,3} (\eta,k)=D_{3,3}(\eta)+kD_{4,1}(\eta)$.) Note that if $k_{i>0}$ is positive, $D_{m,n}$ is defined as in \eref{Dk+}, as we need to use the inequality $1/\mu<1/\hat\mu<\si/\hat\mu$ in \eref{ctoD}. However, a negative $k_{i>0}$ is allowed, and if $k_{i>0}<0$, we must define $D_{m,n}$ as
\be
\label{Dk-}
D_{m, n} \= \sum_{{\rm even} ~i= 0}^n  \sigma^{n-i} M_{m}^{i} \tilde L_{\ell}^{n-i} -\sum_{{\rm odd} ~i= 1}^n   M_{m}^{i} \tilde L_{\ell}^{n-i}  , ~~~{\rm if}~k_{i>0}<0  .
\ee
We emphasize that the two definitions of $D_{m,n}$ are essentially the same if $(\epi\Lambda)^2\gg m^2$.

For a given set of $k_i$, we can again minimize ${D}^{\rm su}_{m,n}(\eta,k)$ with respect to $\eta$ to get
\be
S_{m,n}(k)= \min_\ei D^{\rm su}_{m,n}(\ei,k)  .
\ee
Following the same procedure as for the single $D_{m,n}(\eta)$ case, we can obtain the $D^{\rm su}$ positivity bounds
\bal
\label{lobod}
c_{m,2k}  + \sum_{i\geq 1} k_i c_{m+ i,2k-2i} & > S_{m,2k}(k)~c_{m+k,0}  ,
\\
\label{oddbound52}
c_{m,2k+1}  + \sum_{i\geq 1} k_i c_{m+ i,2k+1-2i}  & >  S_{m,2k+1}(k) \sqrt{c_{m+k,0}c_{m+k+1,0}} ,
\eal
where $k_i$ are constants that can be arbitrarily chosen in order to get the best bounds.

Combing different ${D}^{\rm su}_{m,n}(\eta)$ at the same level ought to give rise to better positivity bounds due to the simple fact that, for a set of functions $f_i(\ei)$, we have $\min_\ei \,\sum_i f_i(\ei)\geq \sum_i \min_\ei\, f_i(\ei)$. Choosing $f_{i}(\ei)$ as $k_i D_{m+ i,n-2i}(\eta)$ with $k_i>0$, this will enhance the inequalities by allowing greater $S_{m,n}$. Note that for an individual bound the effect of this enhancement is not easy to see, as adding a $D_{m+ i,n-2i}(\eta)$ introduces an extra $c_{m+ i,n-2i}$ into the inequality; the enhancement can be seen after one puts all the inequalities together, which is also what we have found empirically, as will be shown in Section \ref{sec:comp}. If one also takes into accounting triple crossing symmetry, as will be discussed shortly in Section \ref{sec:triple}, even for those combinations with $k_i<0$, we will have the enhancement in the sense of ``$\min\sum \geq \sum \min$'' stated above, because with triple crossing symmetry we can get inequalities going the opposite direction for a given $c_{m,n}$.

\subsection{Triple crossing symmetric positivity bounds}
\label{sec:triple}

In Section \ref{sec:nonlinear} and \ref{sec:evenmoreb}, we have only used the $s\leftrightarrow u$ symmetry of a scalar amplitude. But of course the scalar amplitude $B(s,t)$ enjoys the bigger triple crossing symmetry $s\leftrightarrow t \leftrightarrow u$. In this subsection, we will exploit this fact to extract more positivity bounds. We should emphasize that the $\epi\Lambda$ subtracted amplitude $B_{\epi\Lambda}(s,t)$ is in general not triple crossing symmetric because the subtraction from $4m^2$ to $\epi\Lambda$ is only $s\leftrightarrow u$ symmetric. Of course, for the tree level amplitude, $B_{\epi\Lambda}(s,t)$ still has the triple crossing symmetry because for that case the subtraction from $4m^2$ to $(\epi\Lambda)^2$ vanishes. That is, we take $(\epi\Lambda)^2$ to be the scale of the first heavy state that lies outside of EFT: $\Lambda_{\rm th}^2$. We will work with this understanding of the notations in this subsection. Also, since for this scenario we are only interested in the case where a healthy hierarchy of scales exists in the EFT $\Lambda_{\rm th}^2\gg m^2$, we shall take the limit $s,t\gg m^2$, and neglect the differences between the hatted and un-hatted variables, thus having
\be
\hat D_{m,n}=D_{m,n}~ ~~{\rm and} ~~~\tilde L^n_\ell = L^n_\ell  .
\ee

Starting from an $s\leftrightarrow u$ symmetric dispersion relation, the triple crossing symmetry can be exploited by further imposing the $s\leftrightarrow t$ symmetry. In Section \ref{sec:tricrotder}, we have derived the simplest of such $s\leftrightarrow t$ crossing constraints, \eref{crossiden1D}, which states that the average, as defined in \eref{averagedef}, over a polynomial of $\ell$ multiplied by powers of $1/\mu$ must vanish to preserve the $s\leftrightarrow t$ symmetry.
However, a more convenient way to impose the triple crossing symmetry is to start with the $s\leftrightarrow u$ symmetric expansion (\ref{bcwt}), and then the $s\leftrightarrow t$ crossing symmetry can be viewed as the redundancy of the expansion coefficients $c_{m,n}$. To extract these redundancies, we may express the $s\leftrightarrow u$ symmetric coefficients $c_{m,n}$ in terms of some triple symmetric coefficients. A good triple symmetric basis to express $c_{m,n}$ can be provided by $a_{i,j}$ as defined in \eref{adef}. Then one can eliminate $a_{i,j}$ in these relations to obtain relations between $c_{m,n}$. The lowest orders of the explicit relations between $c_{m,n}$ and the independent $c_{m,n}$ can be found in Table \ref{tab:cmn}. The black $c_{m,n}$ in the table can be viewed as independent. Note that the number of $c_{m,n}$ at level $1/\mu^{N-2}$ should be the number of possible $(m,n)$ solutions to equation $2m+n=N$, while the number of independent $c_{m,n}$ at level $1/\mu^{N-2}$ should be the number of possible $(i,j)$ solutions to equation $2i+3j=N$. The difference between them is the number of the constraints which are shown in Table \ref{tab:cmn}. From \eref{cml}, we can see that these constraints on $c_{m,n}$ can be viewed as conditions imposed on the integral average of the polynomials of $\ell$ with the density $\rho _{\ell,\alpha}$, generalizing the condition of \eref{crossiden1D}.

Let us use a simple example to illustrate how this table is obtained. An amplitude may be expressed either $(w,t)$ or $(x,y)$, which are related by $x=w+t^2$, $y=wt$:
\bal
\label{Awt1}
B_{\rm tr}(s,t)&=c_{1,0}w+c_{0,2}t^2+c_{1,1}wt+c_{0,3}t^3+c_{2,0}w^2+c_{1,2}wt^2+c_{0,4}t^4+...\\
B_{\rm tr}(s,t)&=a_{1,0}x+a_{0,1}y+a_{2,0}x^2+...  ,\\
\label{Axy1}
&=a_{1,0}w+a_{1,0}t^2+a_{0,1}wt+a_{2,0}w^2+2a_{2,0}wt^2+a_{2,0}t^4+... \ .
\eal
Matching \eref{Awt1} to \eref{Axy1}, we can get $c_{1,2}=2a_{2,0}=2c_{2,0}$, which is the first redundancy shown in Table \ref{tab:cmn}.

\begin{table}
\small
\begin{center}
\begin{tabular}{|c"ccccccc|p{0.47\columnwidth}|}
\hline
 &$\ei^0$&$\ei^2$&$\ei^4$&$\ei^6$&$\ei^8$&$\ei^{10}$&$\ei^{12}$&~~~~~~$\Gi_{m,n}$ constraints from $s\leftrightarrow t$ symmetry\\
\thickhline
$\mu^{2}$&\textcolor{green}{$c_{0,0}$}& & & & & & & \\
\hline
$1$&$c_{1,0}$&\textcolor{green}{$c_{0,2}$}& & & & & &\\
\hline
$\f1{\mu^{2}}$&$c_{2,0}$&\textcolor{blue}{$c_{1,2}$}&\textcolor{green}{$c_{0,4}$}& & & & & $\textcolor{blue}{c_{1,2}}-2c_{2,0}=0$\\
\hline
$\f1{\mu^{4}}$&$c_{3,0}$&$c_{2,2}$&\textcolor{blue}{$c_{1,4}$}&\textcolor{green}{$c_{0,6}$}& & & &$\textcolor{blue}{c_{1,4}}-3c_{3,0}=0$\\
\hline
$\f1{\mu^{6}}$&$c_{4,0}$&$c_{3,2}$&\textcolor{blue}{$c_{2,4}$}&\textcolor{blue}{$c_{1,6}$}&\textcolor{green}{$c_{0,8}$}& &&$\textcolor{blue}{c_{2,4}}-c_{3,2}-2c_{4,0}=0$,\ $\textcolor{blue}{c_{1,6}}-2\textcolor{blue}{c_{2,4}}-2c_{3,2}-8c_{4,0}=0$\\
\hline
$\f1{\mu^{8}}$&$c_{5,0}$&$c_{4,2}$&\textcolor{blue}{$c_{3,4}$}&\textcolor{blue}{$c_{2,6}$}&\textcolor{blue}{$c_{1,8}$}&\textcolor{green}{$c_{0,10}$}& &$\textcolor{blue}{c_{3,4}}-2c_{4,2}=0$, $\textcolor{blue}{c_{2,6}}+\textcolor{blue}{c_{3,4}}-3c_{4,2}-5c_{5,0}=0$,\ $\textcolor{blue}{c_{1,8}}+4\textcolor{blue}{c_{2,6}}+3\textcolor{blue}{c_{3,4}}-10c_{4,2}-25c_{5,0}=0$\\
\hline
$\f1{\mu^{10}}$&$c_{6,0}$&$c_{5,2}$&$c_{4,4}$&\textcolor{blue}{$c_{3,6}$}&\textcolor{blue}{$c_{2,8}$}&\textcolor{blue}{$c_{1,10}$}&\textcolor{green}{$c_{0,12}$}& $\textcolor{blue}{c_{3,6}} \!-\! 3c_{5,2}\!-\!2c_{6,0}\!=\!0$, $\textcolor{blue}{c_{2,8}}\!+\!3\textcolor{blue}{c_{3,6}}\!-\!10c_{5,2}\!-\!15c_{6,0}\!=\!0$, $\textcolor{blue}{c_{1,10}}+6\textcolor{blue}{c_{2,8}}+12\textcolor{blue}{c_{3,6}}-42c_{5,2}-84c_{6,0}=0$ \\
\hline
\end{tabular}

\vskip 10pt
\begin{tabular}{|c"cccccc|p{0.53\columnwidth}|}
\hline
~ &$\ei^1$&$\ei^3$&$\ei^5$&$\ei^7$&$\ei^9$&$\ei^{11}$& ~~~~~~~$\Gi_{m,n}$ constraints from $s\leftrightarrow t$ symmetry\\
\thickhline
$\mu$&\textcolor{green}{$c_{0,1}$}& & & & & &\\
\hline
$\f1{\mu}$&$c_{1,1}$&\textcolor{green}{$c_{0,3}$}&  & & & &\\
\hline
$\f1{\mu^{3}}$&$c_{2,1}$&\textcolor{blue}{$c_{1,3}$}&\textcolor{green}{$c_{0,5}$} & & & &
$\textcolor{blue}{c_{1,3}}-c_{2,1}=0$\\
\hline
$\f1{\mu^{5}}$&$c_{3,1}$&\textcolor{blue}{$c_{2,3}$}&\textcolor{blue}{$c_{1,5}$}&\textcolor{green}{$c_{0,7}$} & & &$\textcolor{blue}{c_{2,3}}-2c_{3,1}=0$,~$\textcolor{blue}{c_{1,5}}+\textcolor{blue}{c_{2,3}}-3c_{3,1}=0$\\
\hline
$\f1{\mu^{7}}$&$c_{4,1}$&$c_{3,3}$&\textcolor{blue}{$c_{2,5}$}&\textcolor{blue}{$c_{1,7}$}&\textcolor{green}{$c_{0,9}$} & &$\textcolor{blue}{c_{2,5}}-3c_{4,1}=0$,~$\textcolor{blue}{c_{1,7}}+3\textcolor{blue}{c_{2,5}}-10c_{4,1}=0$\\
\hline
$\f1{\mu^{9}}$&$c_{5,1}$&$c_{4,3}$&\textcolor{blue}{$c_{3,5}$}&\textcolor{blue}{$c_{2,7}$}&\textcolor{blue}{$c_{1,9}$}& \textcolor{green}{$c_{0,11}$}& $\textcolor{blue}{c_{3,5}} - c_{4,3}- 2c_{5,1}=0$,~$\textcolor{blue}{c_{2,7}}+2\textcolor{blue}{c_{3,5}}-2c_{4,3}-8c_{5,1}=0$,\ $\textcolor{blue}{c_{1,9}}+5\textcolor{blue}{c_{2,7}}+7\textcolor{blue}{c_{3,5}}-7c_{4,3}-35c_{5,1}=0$ \\
\hline
\end{tabular}
\end{center}
\caption{First constraints on $s\leftrightarrow u$ symmetric coefficients $c_{m,n}$ from triple crossing symmetry.
The label $\ei^n$ and $1/\mu^{2m+n-2}$ can be understoon from \eref{ctoD}. The green $c_{m,n}$ come from the $b(t)$ term in \eref{bcwt}, which can be fixed by the $s\leftrightarrow t$ symmetry: $c_{0,2k}=c_{k,0}$ and $c_{0,2k+1}=0$. The black ones can be chosen as the independent set of $c_{m,n}$ with the full triple crossing symmetry. We have preferred less $t$ derivative coefficients because they are simpler. The blue ones can be expressed as linear combinations of the black ones. We have listed all the constraints up to level $1/\mu^{10}$ and further constraints at least up to order $1/\mu^{22}$ can be obtained by \eref{ctoD}, \eref{GiDD} and recurrence relation (\ref{GiGiGi}).}
\label{tab:cmn}
\end{table}

As we can see in Table \ref{tab:cmn}, these triple symmetric constraints appear as linear combinations of $c_{m,n}$ at the same level of $1/\mu$. Generally, they take the form
\beq
\label{Giconstraint}
\<\frac{\Gi_{m,n}(\ei)}{\mu^{2m+n-2}}\>=c_{m,n}+ \sum_{2k+l=2m+n} \gi_{k,l} c_{k,l} = 0  ,
\eeq
where $\gi_{i,j}$ can be read from the table above. As mentioned above, for a given level of $1/\mu$, there may be several $\Gi_{m,n}$, as shown in Table \ref{tab:cmn}, which is the difference between the numbers of $c_{m,n}$ and $a_{i,j}$ with $(m,n)$ and $(i,j)$ satisfying $2m+n=2i+3j=N$.  In Table \ref{tab:cmn} and \eref{Giconstraint}, we have adapted a convention that $\Gi_{m,n}$ is an $n$-order polynomial of $\ei$ that has a positive coefficient for the highest order term.

From the explicitly obtained constraints, we find that $\Gi_{m,m+1}(\ei)$ satisfies the following simple relations
\bal
\label{GiDD}
\Gi_{m,m+1}(\ei) &= D_{m,m+1}(\ei)-2D_{m+1,m-1}(\ei)   ,
\\
\label{GiGiGi}
\Gi_{m,n}(\ei)&=\Gi_{m+1,n}(\ei)+\Gi_{m,n-1}(\ei)  .
\eal
We have not proven them generically, but they have been explicitly verified up to level $1/\mu^{22}$. It is not difficult to see that general $\Gi_{m,n}(\ei)$ can be generated by $\Gi_{m,m+1}(\ei)$ via the relation
\beq
\label{Gihigh}
\Gi_{m,n}(\ei)=\sum_{i=0}^{n-m-1} \frac{(n-m-1)!}{i!(n-m-1-i)!} \Gi_{n-i-1,n-i}(\ei)   .
\eeq
There is only one such ``generator'' $\Gi_{m,m+1}(\ei)$ at each column in the Table \ref{tab:cmn}.
There are some interesting properties of $\Gi_{m,n}$. For example, there are only $n-1$ constraints at order $\eta^n$: $\Gi_{1,n}$, $\Gi_{2,n}$ ... $\Gi_{n-1,n}$. So the constraints $\Gi_{m,n}$ appear only when $m<n$, our independent $c_{m,n}$ being those with $m \geq n$. Also, $\Gi_{m,n}$ does not have any $\eta^0$ term, {\it i.e.}, $\Gi_{m,n}(\ei=0)=0$.

\subsubsection{The $\bar{D}^{\rm stu}$ positivity bounds}
\label{sec:Dbarstu}

To see how these constraints can be used to derive new positivity bounds, let us consider a linear combination at level $1/\mu^{2m+n-2}$
\be
\bar{D}^{\rm stu}_{m,n}(\eta,k,\ki)=-D^{\rm su}_{m,n}(\eta,k)+\sum_{j\geq 0}\ki_j \Gi_{m'+j,n'-2j}(\ei)  ,
\ee
where $j$ runs over all possible $s\leftrightarrow t$ constraints at level $1/\mu^{2m+n-2}$ and
\be
2m'+n'=2m+n,~~~n’>n,~~~\kappa_0>0   .
\ee
Since $D^{\rm su}_{m,n}(\eta,k)$ is an $n$-th order polynomial of $\eta$ and $\Gi_{m',n'}$ is an $n'$-th order polynomial of $\ei$ that has a positive coefficient for the highest order term,   $\bar{D}^{\rm stu}_{m,n}(\eta,k,\ki)$ must have a minimum with respect to $\eta$: $T^\ki_{m,n}(k)={\rm min}_\ei \bar{D}^{\rm stu}_{m,n}(\eta,k,\ki)$. This leads to a multi-parameter family of positivity bounds, with the choice of different $k_i$ and $\ki_j$, and we can vary $\ki_j$ get the best bounds, which is to take the maximum of $T^\ki_{m,n}(k)$ with respect to $\ki_j$:
\be
T_{m,n}(k)={\rm max}_{\ki}T^\ki_{m,n}(k)={\rm max}_{\ki}{\rm min}_\ei \bar{D}^{\rm stu}_{m,n}(\eta,k,\ki)  ,
\ee
\be
 - \< \frac{D^{\rm su}_{m,n}(\eta,k)}{\mu^{2m+n-2}} \> =
\< \frac{-D^{\rm su}_{m,n}(\eta,k)+\sum_{j\geq 0}\ki_j \Gi_{m'+j,n'-2j}(\ei)}{\mu^{2m+n-2}} \>
> T_{m,n}(k) \< \frac{1}{\mu^{2m+n-2}} \>  .
\ee
Note that $k_i$ and $\ki_j$ are different in that adding a $k_i$ will introduce a $c_{m,n}$ into the positivity bound, while adding $\ki_j$ will not. That is why we can vary $\ki_j$ to maximize $T^\ki_{m,n}(k)$ to get a {\it bonafide} best bound for given $k_i$, but not so to vary $k_i$ to maximize $T^\ki_{m,n}(k)$. With extra parameters coming into play, the bounds with different $k_i$ are often complementary to each other, restricting the parameter space in different directions, as one can see late from the explicit examples of  Fig.~\ref{fig:c41} and~\ref{fig:c52}. If one were to vary $k_i$ to maximize $T^\ki_{m,n}(k)$, it would correspond to restricting the bounds with maximum intersections with the vertical axis in these figures, which does not necessarily give rise to the best bounds.

In practice, our empirical results indicate that adding the $\ki_{j\geq1}$ does not seem to improve the bound. So in the rest of the paper, we will set
\be
\ki_j=0,~~~j\geq 1 ,
\ee
and only need to add the constraint $\Gi_{m',n'}(\ei)$ with the highest $n'$ to $\bar{D}^{\rm stu}_{m,n}(\eta,k,\ki)$:
\beq
\bar{D}^{\rm stu}_{m,n}(\eta,k,\ki) = -D^{\rm su}_{m,n}(\eta,k)+\ki\Gi_{m-l,n+2l}(\ei)     \geq T_{m,n},~l=\left\lfloor\frac{m-n}{3}\right\rfloor+1, ~~~\ki>0  .
\eeq
So we have the positivity bound
\beq
\label{tcsbound}
- \< \frac{D^{\rm su}_{m,n}(\eta,k)}{\mu^{2m+n-2}} \> = \<  \frac{-D^{\rm su}_{m,n}(\eta,k)+\ki\Gi_{m',n'}(\ei)}{\mu^{2m+n-2}} \>
> T_{m,n}(k) \< \frac{1}{\mu^{2m+n-2}} \>  ,
\eeq
which leads to the $\bar{D}^{\rm stu}$ positivity bounds
\bal
\label{upbod}
c_{m,2k}  + \sum_{i\geq 1} k_i c_{m+ i,2k-2i} & < -T_{m,2k}(k)~c_{m+k,0}  ,
\\
\label{boundood34}
c_{m,2k+1}  + \sum_{i\geq 1} k_i c_{m+ i,2k+1-2i}  & <  -T_{m,2k+1}(k) \sqrt{c_{m+k,0}c_{m+k+1,0}}  ,
\eal
where $k_i$ are constants that we can choose to get the best bounds and the bound (\ref{boundood34}) can be obtained when $T_{m,2k+1}(k)<0$. Note that an important feature these bounds are that, compared with the bounds (\ref{lobod}) and (\ref{oddbound52}) in Section \ref{sec:evenmoreb}, because of the minus sign in the front of $D_{m,n}$, the direction of the inequality is reversed for the $c_{m,n}$ term, so these new bounds are complementary to each other, restricting the Wilson coefficients from both directions.

Determining the maximum of the minimum of $\bar{D}^{\rm stu}_{m,n}(\eta,k,\ki)$ is typically easier than it looks. Note that we do not want $\ki$, which is positive, to be too small or large, since taking the limit $\ki\rightarrow0$ or $\ki\rightarrow\infty$ would make the polynomial unbounded below. For fixed $\eta$ and $k_i$, $\bar{D}^{\rm stu}_{m,n}(\eta,k,\ki)$ is a straight line in the $\bar{D}^{\rm stu}_{m,n}$-$\ki$ plot. So it is suffice to evaluate $\bar{D}^{\rm stu}_{m,n}(\eta,k,\ki)$ for the first few $\eta$ and find the maximum of the minimum envelope of different $\bar{D}^{\rm stu}_{m,n}(\eta,k,\ki)$ with $\eta=\ell(\ell+1),~\ell=0,2,4, ...$.

\begin{figure}[h]
\centering
\includegraphics[width=0.5\textwidth]{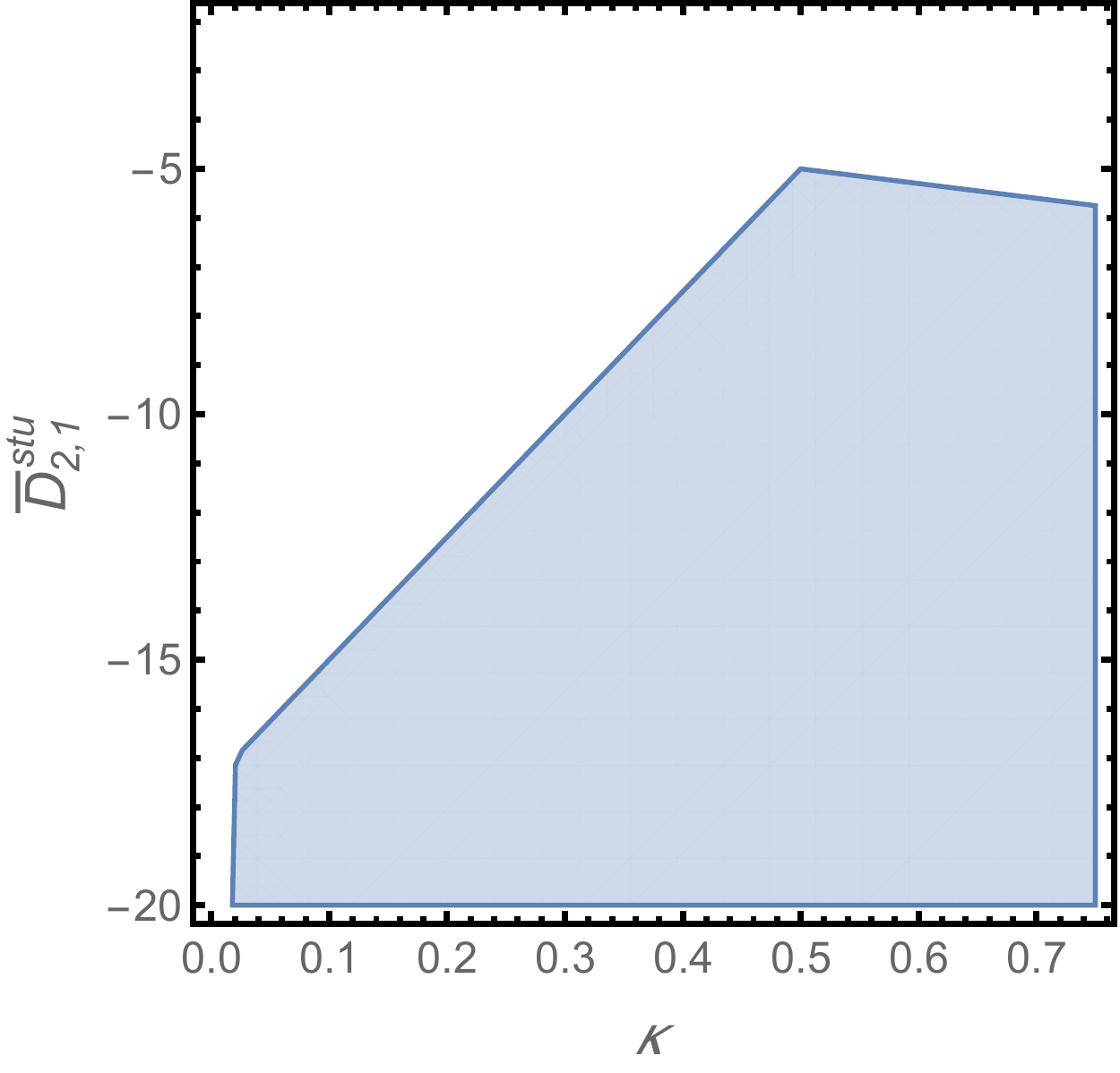}
\caption{Convex hull of $\bar{D}^{\rm stu}_{2,1}(\eta,\ki)$ in 4D for different $\eta=\ell(\ell+1),~\ell=0,2,4, ...$. This convex hull gives $T^\ki_{2,1}={\rm min}_\ki \bar{D}^{\rm stu}_{2,1}(\eta,\ki)$. The maximum of $T^\ki_{2,1}$ over positive $\ki$ gives $T_{2,1}$ in the positivity bound (\ref{boundood34}), which is at the intersection point of line $\bar{D}^{\rm stu}_{2,1}(12,\ki)$ and line $\bar{D}^{\rm stu}_{2,1}(20,\ki)$.}
\label{fig:Dbar21}
\end{figure}

It is easiest to understand the procedure by a couple of examples. First, suppose that we want to find $T_{2,1}={\rm max}_\ki {\rm min}_\ei \bar{D}^{\rm stu}_{2,1}(\eta,\ki)$ in 4D. At that level, there is only a single $D_{m,n}(\ei)$, so we only need to consider $\bar{D}^{\rm stu}_{2,1}(\ei,\ki)=-D_{2,1}(\ei)+\ki\Gi_{1,3}(\ei)$. A straightforward computation gives
\beq
\label{examplestu1}
\bar{D}^{\rm stu}_{2,1}(\ei,\ki)=-D_{2,1}+\ki\Gi_{1,3}=-\ei+\frac{5}{2}
+\ki \(\frac{1}{36}\ei^3-\frac{43}{72}\ei^2+\frac{25}{12}\ei\)  .
\eeq
Evaluating this expression for $\eta=\ell(\ell+\ai)$ with $\ell=0,2,4,...$, we have
\bal
\bar{D}^{\rm stu}_{2,1}(0,\ki)=\frac{5}{2},~
%\bar{D}^{\rm stu}_{2,1}(2,\ki)= 2\ki+\frac{1}{2},~
&\bar{D}^{\rm stu}_{2,1}(6,\ki)= -3\ki-\frac{7}{2},~
%\bar{D}^{\rm stu}_{2,1}(12,\ki)= -13\ki-\frac{19}{2} ,\\
\bar{D}^{\rm stu}_{2,1}(20,\ki)= 25\ki-\frac{35}{2},~
%\bar{D}^{\rm stu}_{2,1}(30,\ki)&= 275\ki-\frac{55}{2},~
\bar{D}^{\rm stu}_{2,1}(42,\ki)= 1092\ki-\frac{79}{2},...\ .
\eal
We then plot all these lines on the $\bar{D}^{\rm stu}_{m,n}$-$\ki$ plot and find the envelope of the minimum of these lines; see Fig.~\ref{fig:Dbar21}. The maximum of this envelope is $T_{2,1}$. For this example, we have $T_{2,1}=-5$, which is at the intersection point of line $\bar{D}^{\rm stu}_{2,1}(12,\ki)$ and line $\bar{D}^{\rm stu}_{2,1}(20,\ki)$ at $\ki=1/2$. By \eref{boundood34}, we have a bound
\be
c_{2,1}<5\sqrt{c_{2,0}~c_{3,0}}  .
\ee

Now, for the example of $D_{3,3}(\ei)$ in 4D, we have $D^{\rm su}_{3,2}(\ei,k) = D_{3,3}(\ei)-k D_{4,1}(\ei)$ and $\bar{D}^{\rm stu}_{3,2}(\ei,k,\ki) = -D_{3,3}(\ei)+k D_{4,1}(\ei)+\ki \Gi_{2,5}(\ei)$, where we have dropped the constraints $\Gi_{1,3}(\eta)$, as it does not improve the bound. If we choose $k=16$, then $-D_{3,3}+16D_{4,1}+\frac{88}{139}\Gi_{2,5} \geq -57$, so the maximum of the minimum envelope of different $\bar{D}^{\rm stu}_{m,n}(\eta,k,\ki)$ is $T_{3,3}(16)=-57$. This gives rise to a bound: $c_{3,3}-16c_{4,1} < 57 \<{1}/{\mu^7}\> < 57 \sqrt{c_{4,0}c_{5,0}}$. On the other hand, if we choose $k=1$, then $T_{3,3}(1)= -{2709}/{5}$, which gives rise to a complementary bound: $c_{3,3}-c_{4,1} < 2709\sqrt{c_{4,0}c_{5,0}}/5$.

\subsubsection{The $D^{\rm stu}$ positivity bounds}
\label{sec:Dstu}

Of course, we can also use the $s\leftrightarrow t$ constraints to upgrade the $D^{\rm su}$ bounds. At level $1/\mu^{2m+n-2}$, we can consider
\be
{D}^{\rm stu}_{m,n}(\eta,k,\ki)=D^{\rm su}_{m,n}(\eta,k)+\sum_{j\geq 0}\ki_j \Gi_{m'+j,n'-2j}(\ei)  ,
\ee
where $j$ runs over all possible $s\leftrightarrow t$ constraints at level $1/\mu^{2m+n-2}$ and
\be
2m'+n'=2m+n,~~~n'<n  .
\ee
Note that here we require $n'<n$ and we do not require $\ki_0>0$. Since $D^{\rm su}_{m,n}(\eta,k)$ is an $n$-th order polynomial of $\eta$ that has a positive coefficient for the highest order term, ${D}^{\rm stu}_{m,n}(\eta,k,\ki)$ must have a minimum with respect to $\eta$: $U^\ki_{m,n}(k)={\rm min}_\ei {D}^{\rm stu}_{m,n}(\eta,k,\ki)$, and we can vary $\ki_j$ get the best bound:
\be
  \< \frac{D^{\rm su}_{m,n}(\eta,k)}{\mu^{2m+n-2}} \> =
\< \frac{D^{\rm su}_{m,n}(\eta,k)+\sum_{j\geq 0}\ki_j \Gi_{m'+j,n'-2j}(\ei)}{\mu^{2m+n-2}} \>
> U_{m,n}(k) \< \frac{1}{\mu^{2m+n-2}} \>   ,
\ee
where
\be
U_{m,n}(k)=\max_{\ki}\min_\ei {D}^{\rm stu}_{m,n}(\eta,k,\ki) .
\ee
In practice, we can also set $\ki_{j\geq 1}=0$.  This gives the $D^{\rm stu}$ positivity bounds
\bal
\label{Dstulobod}
c_{m,2k}  + \sum_{i\geq 1} k_i c_{m+ i,2k-2i} & > U_{m,2k}(k) c_{m+k,0}  ,
\\
\label{Dstuoddbound52}
c_{m,2k+1}  + \sum_{i\geq 1} k_i c_{m+ i,2k+1-2i}  & >  U_{m,2k+1}(k) \sqrt{c_{m+k,0}c_{m+k+1,0}}  .
\eal
where $k_i$ are constants that can be chosen at will to get the best bounds.

Note that the $D^{\rm stu}$ and $\bar{D}^{\rm stu}$ bounds constrain the linear combinations of all $c_{m,n}$ at the same level $2m+n={fixed}$ in opposite directions. For a given set of $k_i$, $U_{m,n}(m,n)$ and $T_{m.n}(k)$ are generically different, so the $D^{\rm stu}$ and $\bar{D}^{\rm stu}$ bounds for a given set of $k_i$ sandwich out a stripe of space in the parameter space spanned by all $c_{m,n}$ at a given level. Therefore, with different choices of $k_i$, the positivity bounds always carve out an enclosed region for the $c_{m,n}$ parameter space at a given level, and this region is always convex.

\begin{table}[h]
\begin{center}
\begin{tabular}{|c"c|c|c|}
\hline
$(m,n)$   & ${D}^{\rm stu}_{m,n}$ bound   & $\bar{D}^{\rm stu}_{m,n}$ bound \\
\thickhline
$(1,1)$   &  $c_{1,1}>  -\frac{3}{2}\sqrt{c_{1,0}c_{2,0}}$  & $c_{1,1} < 8 \sqrt{c_{1,0}c_{2,0}}$ \\
\hline
$(2,1)$   &    $c_{2,1}  > -\frac{5}{2}\sqrt{c_{2,0}c_{3,0}}$     & $c_{2,1} <  5\sqrt{c_{2,0}c_{3,0}}$  \\
\hline
$(2,2)$  &        $c_{2,2} >  -\frac{9}{2}c_{3,0}$           &   $c_{2,2} <  \frac{93}{2}c_{3,0}$       \\
 \hline
$(3,1)$  &          $c_{3,1} >  -\frac{7}{2}\sqrt{c_{3,0}c_{4,0}}$        &    $c_{3,1} <  \frac{35}{2}\sqrt{c_{3,0}c_{4,0}}$    \\
\hline
$(3,2)$   &         $c_{3,2} >  -7c_{4,0}$             &  $c_{3,2} <  \frac{227}{4}c_{4,0}$  \\
 \hline
 $(3,3)$   &  $c_{3,3}+\frac{3}{4}c_{4,1}>-\frac{147}{8}\sqrt{c_{4,0}c_{5,0}}$,  &  $c_{3,3}-\frac{1982}{141}c_{4,1}<\frac{2268}{47}\sqrt{c_{4,0}c_{5,0}}$  \\
               &  $c_{3,3}-8c_{4,1}>-154\sqrt{c_{4,0}c_{5,0}}$,   & \\
               &  $c_{3,3}-\frac{481}{12}c_{4,1}>-\frac{7777}{8}\sqrt{c_{4,0}c_{5,0}}$, &  \\
               &  $c_{3,3}-104c_{4,1}>-3369\sqrt{c_{4,0}c_{5,0}}$   &  \\
 \hline
$(4,2)$ &       $c_{4,2} >  -\frac{17}{2}c_{5,0}$                &  $c_{4,2} <  \frac{1382}{5}c_{5,0}$     \\
 \hline
$(4,3)$   &  $c_{4,3}+\frac{3}{4}c_{5,1}>-\frac{253}{8}\sqrt{c_{5,0}c_{6,0}}$,  &  $c_{4,3}-\frac{869}{46}c_{5,1}<\frac{7029}{92}\sqrt{c_{5,0}c_{6,0}}$  \\
               &  $c_{4,3}-15c_{5,1}>-260\sqrt{c_{5,0}c_{6,0}}$,   & \\
               &  $c_{4,3}-\frac{277}{2}c_{5,1}>-\frac{19071}{4}\sqrt{c_{5,0}c_{6,0}}$ &  \\
 \hline
 $(4,4)$ & $c_{4,4}+\frac{25}{24}c_{5,2}>-\frac{375}{8}c_{6,0}$,  & $c_{4,4}-15c_{5,2}<\frac{195}{2}c_{6,0}$,\\
             & $c_{4,4}-\frac{53}{5}c_{5,2}>-\frac{24567}{10}c_{6,0}$, & $c_{4,4}-\frac{386865}{22132}c_{5,2}<\frac{1880775}{22132}c_{6,0}$,\\
             & $c_{4,4}-\frac{4370}{79}c_{5,2}>-\frac{6619665}{158}c_{6,0}$ & \\
 \hline
\end{tabular}
\end{center}
\caption{Explicit triple crossing positivity bounds in 4D up to level $1/\mu^{10}$. $c_{m,n}$ are the expansion coefficients of the pole subtracted amplitude in terms of $w$ and $t$ (see \eref{bcwt}). These bounds are valid for weakly coupled UV completions where we choose $\si=1$ (see \eref{sidef}).}
\label{tab:firstbounds}
\end{table}

\subsubsection{Frist few triple crossing bounds}
\label{sec:tripleEg}

The ${D}^{\rm stu}$ and $\bar{D}^{\rm stu}$ positivity bounds contain arbitrary constants $k_i$ and $U_{m,n}(k)$ and $T_{m,n}(k)$ that should be computed for a given set of $k_i$. While an optimal set of $k_i$ and $U_{m,n}(k)$ and $T_{m,n}(k)$ can be straightforwardly computed to give the best bounds, it is nevertheless a tedious task. In an actual application of EFT, one usually restricts to the lowest few orders. For an easy reference, here we list the best ${D}^{\rm stu}$ and $\bar{D}^{\rm stu}$ positivity bounds of the lowest few orders in 4D, up to level $1/\mu^{10}$.

\begin{figure}[h]
\centering
\includegraphics[width=0.5\textwidth]{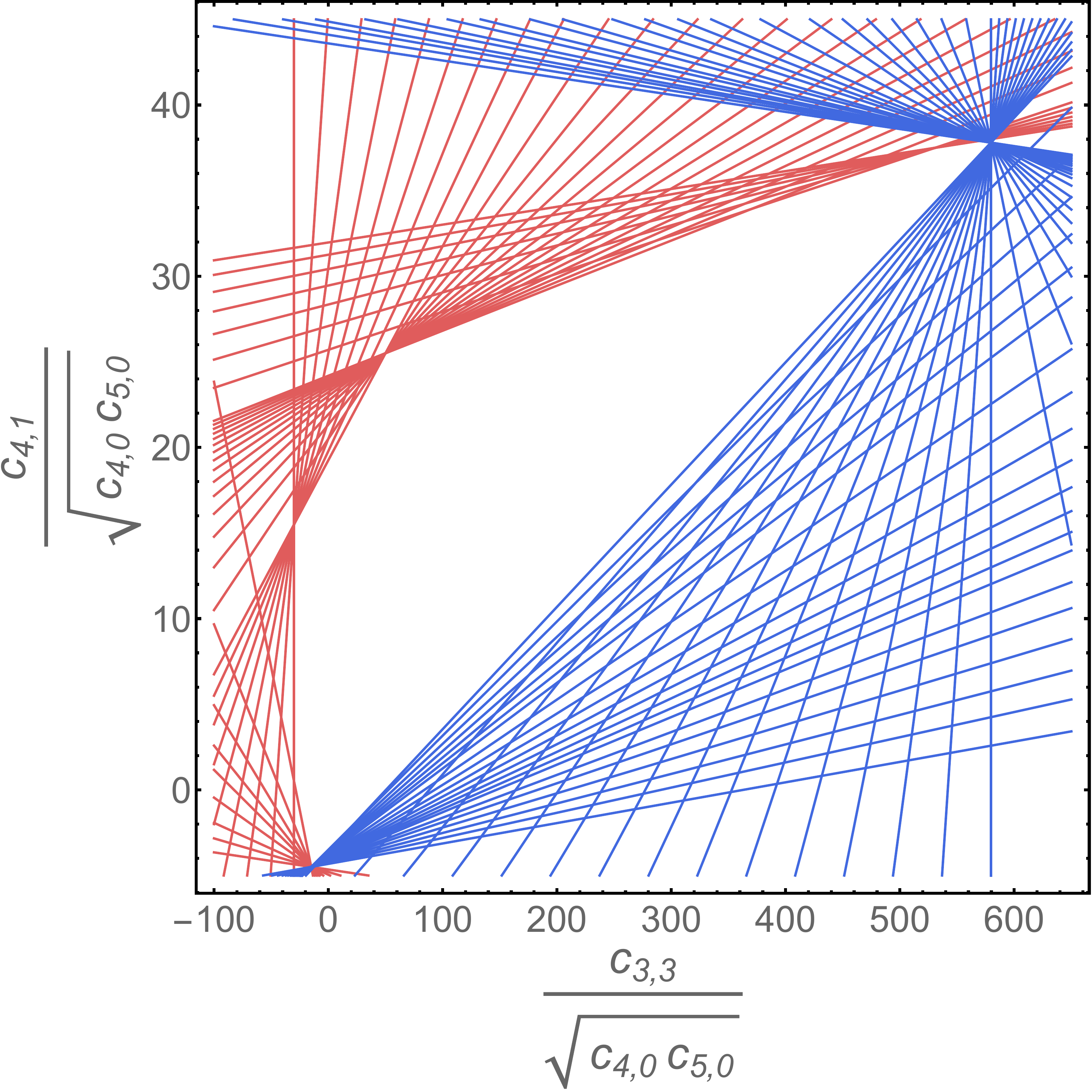}
\caption{Positivity bounds on $c_{3,3}/\sqrt{c_{4,0}c_{5,0}}$ and $c_{4,1}/\sqrt{c_{4,0}c_{5,0}}$. The red (blue) lines are the ${D}^{\rm stu}_{3,3}$ ($\bar{D}^{\rm stu}_{3,3}$) bounds with different choices of $k$. The enclosed region pentagon is the region allowed by the optimal positivity bounds. Non-optimal positivity bounds are also plotted with equal interval choices of $k$.}
\label{fig:c41}
\end{figure}

\begin{figure}[h]
\centering
\includegraphics[width=0.5\textwidth]{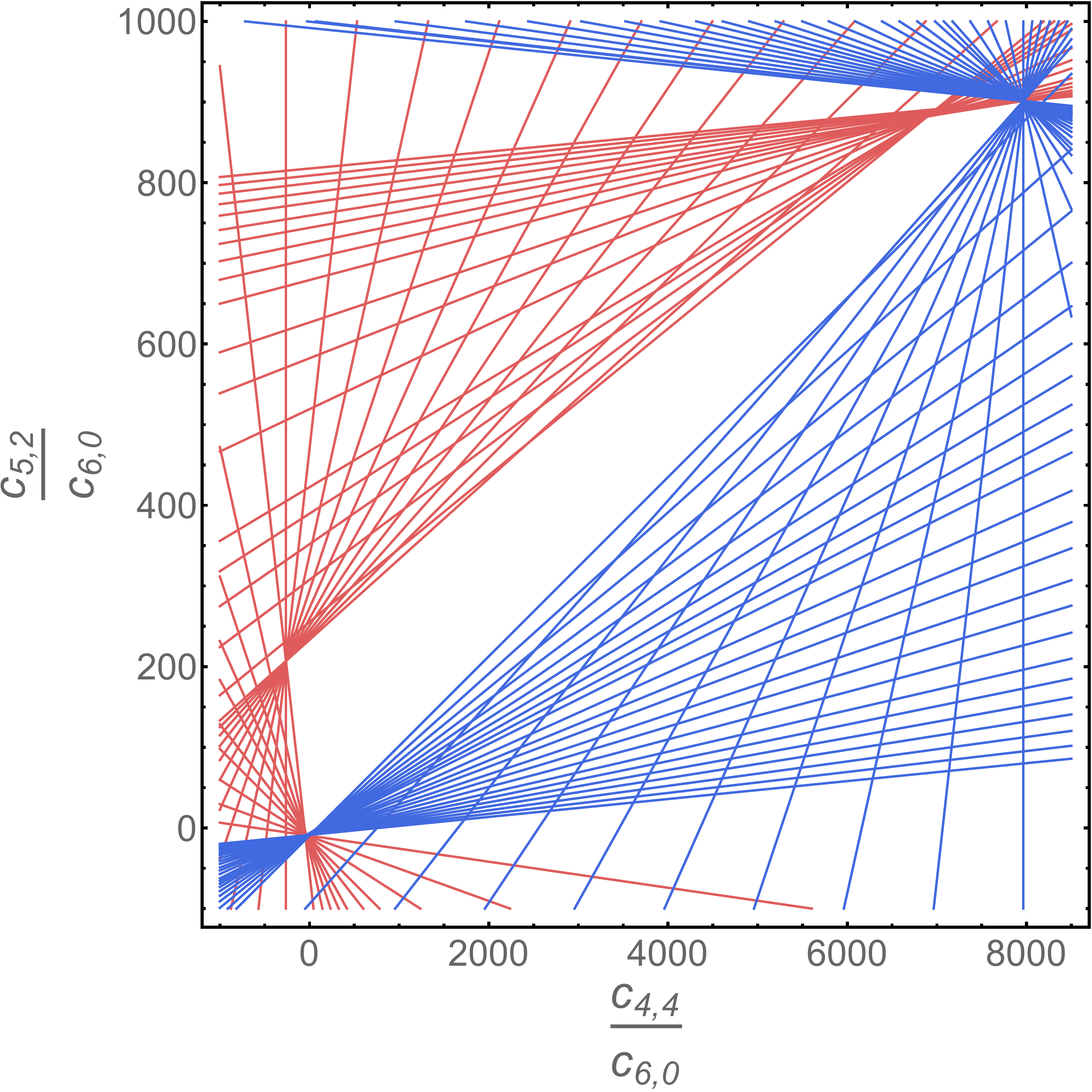}
\caption{Positivity bounds on $c_{4,4}/c_{6,0}$ and $c_{5,2}/c_{6,0}$. The red (blue) lines are the ${D}^{\rm stu}_{4,4}$ ($\bar{D}^{\rm stu}_{4,4}$) bounds with different choices of $k$. The enclosed region hexagon is the region allowed by the optimal positivity bounds.}
\label{fig:c52}
\end{figure}

First, we have two $t$-derivative bounds on $c_{1,1}$ (see \eref{t1boounc}): $-\frac{3}{2}\sqrt{c_{1,0}c_{2,0}} <  c_{1,1} < 8 \sqrt{c_{1,0}c_{2,0}}$, which are derived in Section \ref{sec:tricrotder} but not obvious from the generic formalism in this section. In fact, every $c_{m,n}$ with $n \neq 0$ can be bounded by $c_{n',0}$ in the generic formalism except $c_{1,1}$. See Table \ref{tab:firstbounds} for all  the ${D}^{\rm stu}$ and $\bar{D}^{\rm stu}$ positivity bounds in 4D up to level $1/\mu^{10}$.
The optimal set of $k_i$, $U_{m,n}$ and $T_{m,n}$ can be read from the bounds directly. For example, for the $D^{\rm stu}_{3,3}$ and $\bar{D}^{\rm stu}_{3,3}$ bounds, the optimal choice is $U_{3,3}(3/4)=-147/8$, $U_{3,3}(-8)=-154$, $U_{3,3}(-481/12)=-7777/8$, $U_{3,3}(-104)=-3369$ and $T_{3,3}(-1982/41)=-2268/47$. If we plot the bounds in the $c_{4,1}$-$c_{3,3}$ plane in the units of $\sqrt{c_{4,0}c_{5,0}}$, we get a pentagon: the enclosed region in Fig \ref{fig:c41}. See also Fig \ref{fig:c52} for the bounds on the $c_{4,4}$-$c_{5,2}$ plane. As one can see in Fig \ref{fig:c41} and Fig \ref{fig:c52}, we always get an enclosed region from the $D^{\rm stu}$ and $\bar{D}^{\rm stu}$ bounds for all $c_{m,n}$ parameters at a given level.

\section{Comparison of the different positivity bounds}
\label{sec:comp}

In this section, we shall compare the different positivity bounds derived here and also in \cite{deRham:2017avq} up to level $1/\mu^4$ in 4D. We will see that the various new positivity bounds are generically complementary to each other and also overlap with the $Y$ bounds. Combining the various bounds, the constraints on the parameter space  can be significantly improved. We shall write the bounds in terms of $\tilde a_{i,j}$, which can be translated from $c_{m,n}$ by \eref{cmntoaij} and \eref{adef}. We will take the limit of $\epi=1$ and $\sigma=1$.

Here, up to level $1/\mu^{4}$ means that we should include all the bounds containing $\tilde a_{i,j}$ with $(i,j)$ satisfying $2i+3j\leq 6$, in which case there are 6 parameters in the triple symmetric amplitude: $\tilde a_{1,0}$, $\tilde a_{2,0}$, $\tilde a_{3,0}$, $\tilde a_{0,1}$, $\tilde a_{1,1}$ and $\tilde a_{0,2}$. All the positivity bounds up to level $1/\mu^{4}$ are as follows:

\begin{itemize}
\item The $Y$ bounds \cite{deRham:2017avq}:
\be
\tilde a_{1,0} > 0, ~\tilde a_{2,0} > 0,~ \tilde a_{3,0} > 0,~ \tilde a_{0,1}+\frac{3}{2}\tilde a_{1,0}> 0, ~ \tilde a_{1,1}+\frac{5}{2 }\tilde a_{2,0}> 0 ,
\ee
\be
\tilde a_{0,2}-\frac{3}{2}\tilde a_{3,0}+\frac{5}{2 }\(\tilde a_{1,1}+\frac{5}{2 }\tilde a_{2,0}\)> 0, ~ 2\tilde a_{0,2}-3\tilde a_{3,0}+\frac{3}{2 }\[-\tilde a_{1,1}+\frac{9}{4} \(\tilde a_{0,1}+\frac{3}{2 }\tilde a_{1,0}\)\] > 0
\ee
\be
\label{-a11}
-\tilde a_{1,1}+\frac{9}{4}\(\tilde a_{0,1}+\frac{3}{2 }\tilde a_{1,0}\) > 0.
\ee

\item The nonlinear $PQ$ bounds:
\be
\tilde a_{1,0} \phantom{.} \tilde a_{3,0} > (\tilde a_{2,0})^2   .
\ee

\item The linear $P>0$ bounds: same as the $Y$ bounds except replacing \eref{-a11} with
\be
-\tilde a_{1,1}+\frac{9}{4\epi^4}(\tilde a_{0,1}+\frac{3}{2 }\tilde a_{1,0})+\frac{5}{2}  \tilde a_{2,0}-3  \tilde a_{3,0} > 0.
\ee

\item The linear $P>Q$ bounds:

\be
\tilde a_{1,0}> \tilde a_{2,0}>  \tilde a_{3,0},~ \tilde a_{0,1}+\frac{3}{2 }\tilde a_{2,0}>  \tilde a_{1,1}+\frac{5 }{2}\tilde a_{3,0}.
\ee

\item The $D^{\rm stu}$ (lower) and $\bar{D}^{\rm stu}$ (upper) bounds:

\be
-\frac{3}{2} \sqrt{\tilde a_{1,0}\tilde a_{2,0}} < \tilde a_{0,1}< 8 \sqrt{\tilde a_{1,0}\tilde a_{2,0}},~
-\frac{5}{2} \sqrt{\tilde a_{2,0}\tilde a_{3,0}} < \tilde a_{1,1}< 5 \sqrt{\tilde a_{2,0}\tilde a_{3,0}},
\ee
\be
-\frac{15}{2} \tilde a_{3,0} < \tilde a_{0,2}< \frac{87}{2} \tilde a_{3,0}  .
\ee

\end{itemize}

\begin{figure}[h]
\centering
\includegraphics[width=0.5\textwidth]{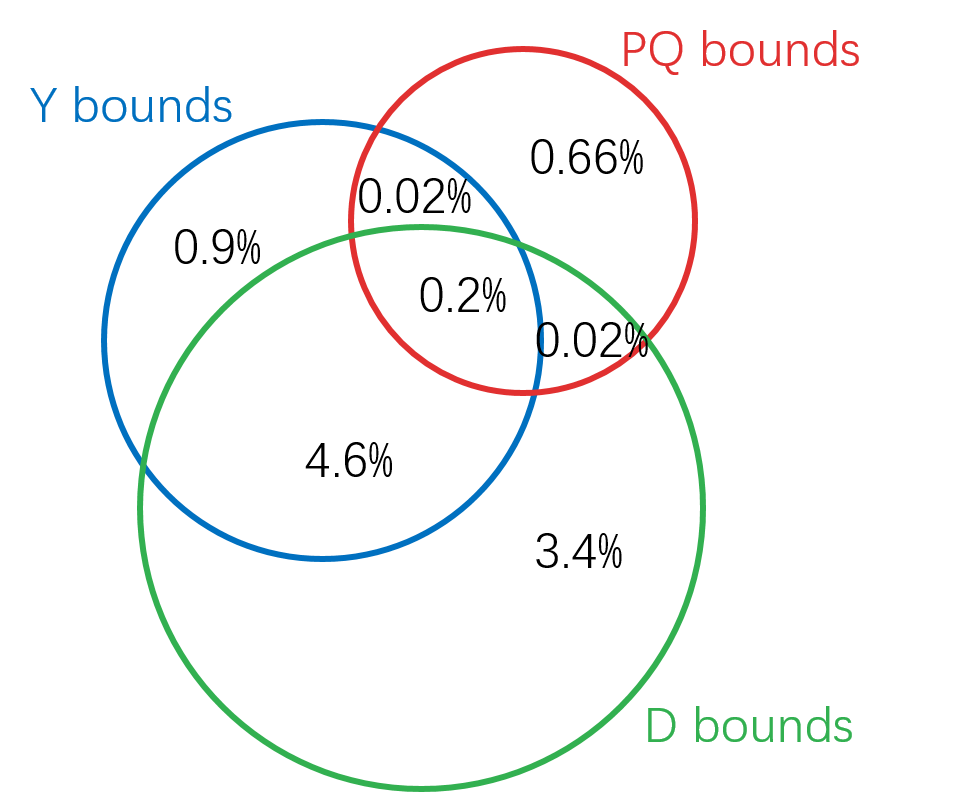}
\caption{Comparison of the $Y$ bounds, $PQ$ bounds and $D$ bounds. The total parameter space is a 6-dimensional sphere. The percentages denoted in the corresponding areas are the percentages of the total solid angle. The blue disk schematically represents the solid angle of the parameter space that satisfies the $Y$ bounds. The red disk represents the solid angle that satisfies the 3 types of $PQ$ bounds, excluding the bounds that are already in the $Y$ bounds. The green circle represents the solid angle that satisfies the $D$ bounds (plus $a_{n,0}>0$, because of the square roots in the $D$ bounds). We assume that the EFT is valid up to the cutoff, so settting $\epi=1$ and $\sigma=1$.}
\label{fig:YPQD}
\end{figure}

\begin{figure}[h]
\centering
\includegraphics[width=0.5\textwidth]{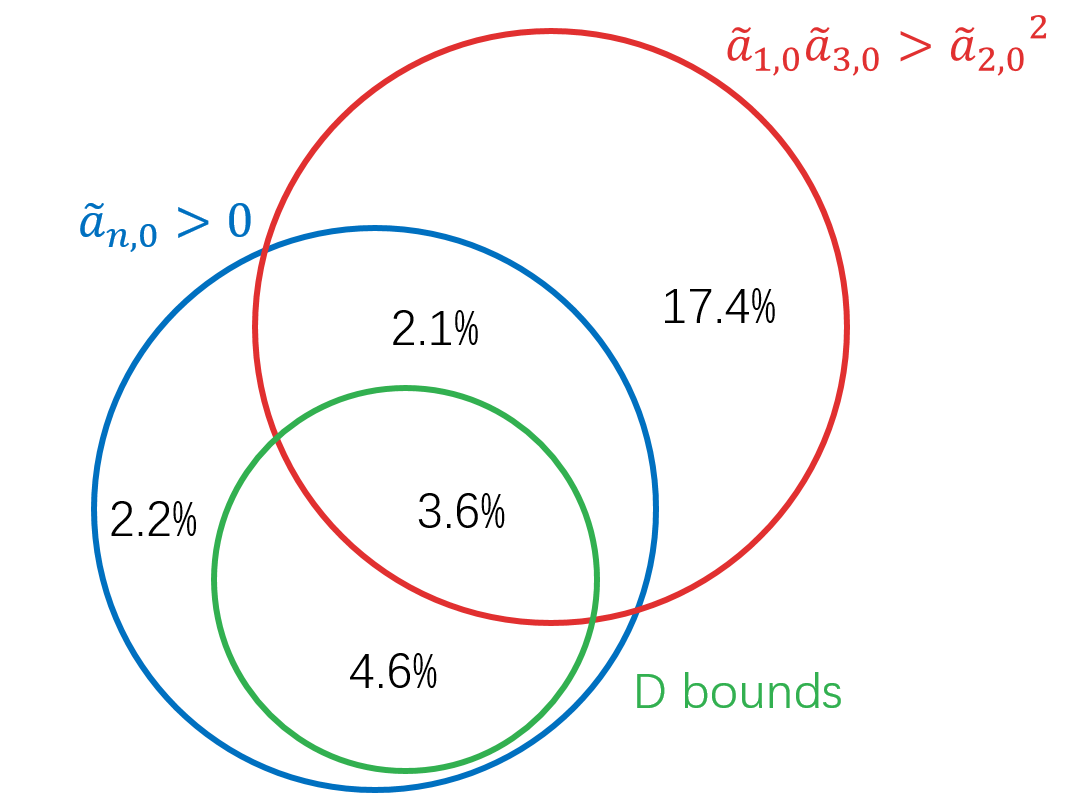}
\caption{Comparison of the $D^{\rm stu}$ and $\bar D^{\rm stu}$ bounds and the nonlinear $PQ$ bounds.}
\label{fig:aD}
\end{figure}

\begin{figure}[h]
\centering
\includegraphics[width=0.5\textwidth]{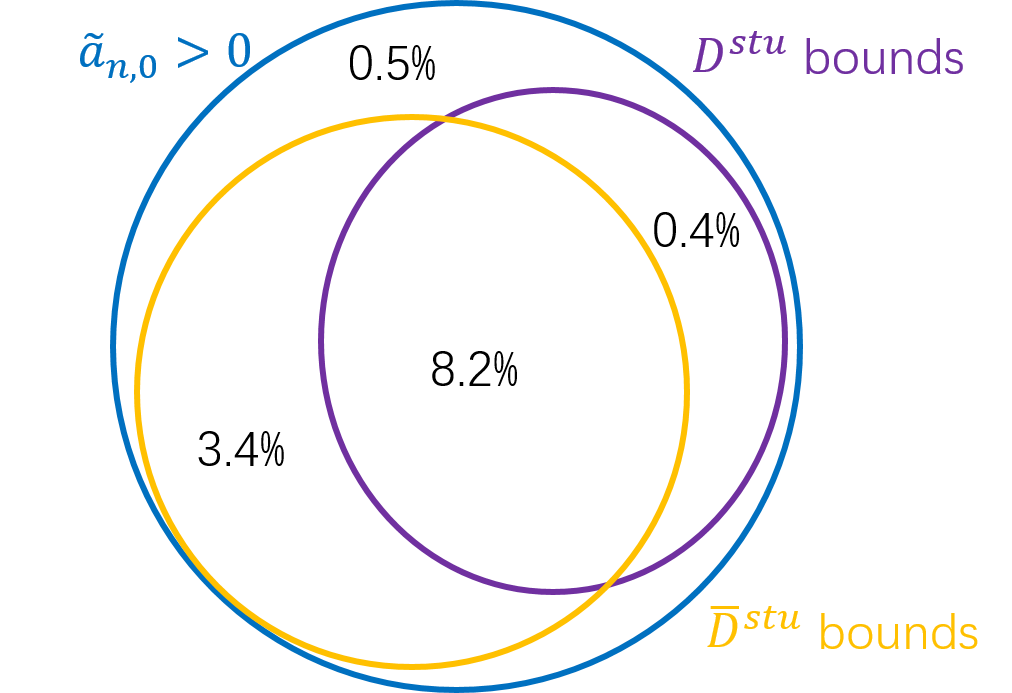}
\caption{Comparison of the $D^{\rm stu}$ bounds and the $\bar{D}^{\rm stu}$ bounds.}
\label{fig:DDbar}
\end{figure}

\begin{figure}[h]
\centering
\includegraphics[width=0.5\textwidth]{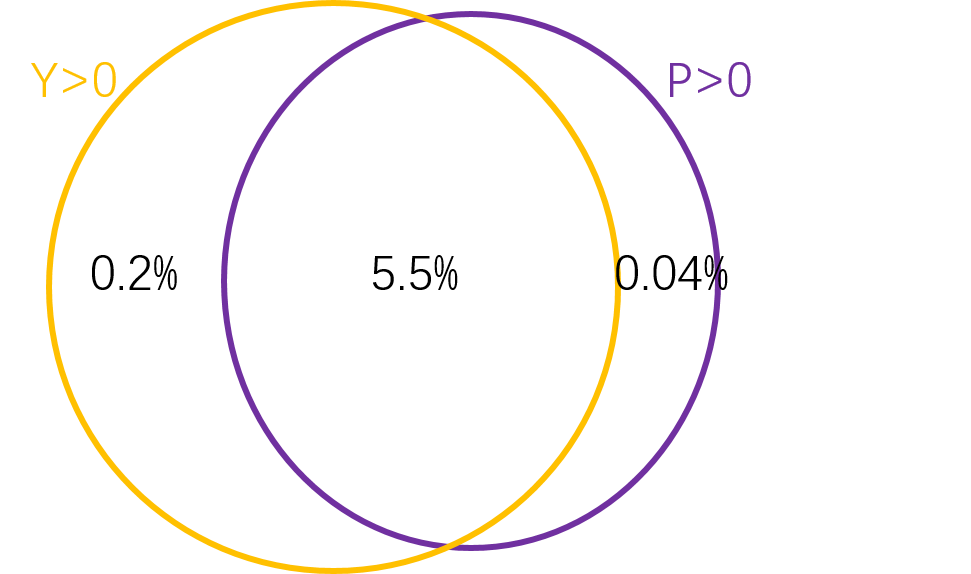}
\caption{Comparison of the $Y$ bounds and the linear $P>0$ bounds. These two sets of bounds are very similar.}
\label{fig:YP}
\end{figure}

Note that all the positivity inequalities above are homogeneous in $a_{i,j}$, so the bounds are invariant under a global scaling of all $a_{i,j}$.
This means that the parameter space allowed by the positivity bounds corresponds to a solid angle on a 6-dimensional sphere ({\it i.e.,} a convex cone). We shall compute the sizes of the solid angles allowed by the various bounds, by statistically sampling the 6-dimensional sphere and counting the percentages of points falling in the various bounds; See Fig \ref{fig:YPQD},  \ref{fig:aD},  \ref{fig:DDbar} and \ref{fig:YP} for schematical portraits of the comparisons (Venn diagrams). As shown in Fig \ref{fig:YPQD}, combining all the available bounds, only $0.2\%$ of the total parameter space of the Wilson coefficients is consistent with the fundamental properties of analytical S-matrix, compared with the previous $Y$ bounds for which $5.7\%$ of the total space is allowed.

As we mentioned at the end of the Introduction, Ref \cite{Caron-Huot:2020cmc} appeared soon after our paper was posted on arXiv, and has also obtained new positivity bounds on a single scalar EFT utilizing full crossing symmetry. As the main idea and methods of Ref \cite{Caron-Huot:2020cmc} are very similar to ours, let us check the consistency between our results. Our full crossing symmetric constraints $\Gamma_{m,n}(\eta)$ (see \eref{Giconstraint}) are dubbed ``null constraints''   $n_i({\cal J}^2)$ in Ref \cite{Caron-Huot:2020cmc}. We have explicitly checked the first few null constraints in Eq (3.29) of \cite{Caron-Huot:2020cmc}), and, as expected, found that they exactly agree with our $\Gamma_{m,n}(\eta)$ constraints. Ref \cite{Caron-Huot:2020cmc} has also explicitly computed the numerical bounds on the first few individual Wilson coefficients, in their notation, $\tilde{g}_{k}^{(p)}=g_{k}^{(p)} M^{2 k-4} / g_{2}$ in their Table 3, up to order $\tilde{g}_{10}^{(p)}$. We can easily obtain these bounds from the bounds in our Table \ref{tab:firstbounds} and additionally the bounds $c_{1,0}>(\epi\Lambda)^2c_{2,0}>...>(\epi\Lambda)^{2m-2}c_{m,0}>0$. The correspondences between our $c_{m,n}$ and their $g_{k}^{(p)}$ are: $c_{1,0} = 2 g_2, c_{1, 1} = -g_3, c_{2, 0} = 4 g_4, c_{2, 1} = -2 g_5, c_{3, 0} = 8 g_6, c_{2, 2} = 24 g_6 + g'_6, c_{3, 1} = -4 g_7, ...$, and note that their $M$ is just our $\epi\Lambda$. We then find that our bounds on $\tilde g_4, \tilde g_6, \tilde g_8 ,\tilde g_{10},...$ are the same as those of Ref \cite{Caron-Huot:2020cmc}, while for others we have $-16< \tilde g_3<3, -5<\tilde g_5 <5/2, -15 < \tilde g'_6<87, -35/4< \tilde g_7<7/4,...$, which are slightly weaker on one side or both sides than those of Ref \cite{Caron-Huot:2020cmc}. This is, of course, expected as they use the semi-definite programing technique to combine crossing symmetric constraints across different levels (or different $n$'s in $1/\mu^n$) to numerically obtain the strongest bounds, while we have only used all available crossing symmetric constraints at the same level, which can be evaluated analytically.

\section{New constraints on chiral perturbation theory}
\label{sec:chpt}

In the previous section, we have compared the effectiveness of the newly obtained $PQ$ and $D$ bounds and the previous $Y$ bounds on constraining the lowest orders of Wilson coefficients in a generic EFT. We have seen that the new bounds give rise to significant improvements in terms of volume statistics. In this section, we shall see the effectiveness the new bounds in another concrete example, SU(2) chiral perturbation theory.

Chiral perturbation theory is often considered as an exemplary EFT, a popular test ground for many new ideas. Indeed, it is the first EFT to be considered following Weinberg's seminal work \cite{Weinberg:1966kf}. It is used in hadron physics to describe the pion interactions, as the low energy EFT of strong coupled quantum chromodynamics, and may also be relevant to other scenarios. The $Y$ positive bounds have recently been applied to SU(2) chiral perturbation theory \cite{Wang:2020jxr}, and it is found that the $Y$ bounds can impose strict bounds on the Wilson coefficients (low energy constants as they are often called there), which improve the older positivity bounds by \cite{Manohar:2008tc}.

The $\pi\pi$ scattering amplitudes in SU(2) chiral perturbation theory has been computed up to $\mc{O}(p^6)$ \cite{Bijnens:1995yn}. We shall apply the $PQ$ and $D^{\rm su}$ bounds to the $\pi^0\pi^0$ loop amplitude. We are not applying the triple crossing $D^{\rm stu}$ and $\bar{D}^{\rm  stu}$ bounds here, because we often consider strongly coupled UV completions for chiral perturbation theory such as in quantum chromodynamics. Also, even if the UV theory is weakly coupled, the tree level amplitude, for which the $D^{\rm stu}$ and $\bar{D}^{\rm  stu}$ bounds can be used, is completely fixed by the symmetries up to an overall constant \cite{Weinberg:1966kf}, which for $\pi^0\pi^0\to \pi^0\pi^0$ is given by
\be
T^{\rm Weinberg}_{\pi^0\pi^0} =  \f{M_\pi^2}{F^2_\pi} (s+t+u-3) ,
\ee
where $M_\pi$ is the pion mass and $F_\pi$ is the pion decay constant, and following the notation of \cite{Bijnens:1995yn}, we define the Mandelstam variables in the units of $M_\pi^2$ such that in this section $s,t,u$ are dimensionless and $s+t+u=4$. This tree amplitude does not contribute to positivity bounds up on the twice subtraction to make use of the Froissart-Martin bound.  Up to one loop, the $\pi^0\pi^0$ amplitude is given by
\be
T_{\pi^0\pi^0} = A(s, t, u) +A(t, s, u)  +A(u, t, s)  ,
\ee
with
\begin{align}
\label{Astupi}
 A(s, t, u)&=   \f{M_\pi^2}{F^2_\pi}  [s-1] +  \f{M_\pi^4}{F^4_\pi} \left[b_{1}+b_{2} s+b_{3} s^{2}+b_{4}(t-u)^{2}\right]
\nn
&~~~ +  \f{M_\pi^4}{F^4_\pi} \left[F^{(1)}(s)+G^{(1)}(s, t)+G^{(1)}(s, u)\right]   ,
\end{align}
where $b_i$ are combinations of the low energy constants and $F^{(1)}(s)$ and $G^{(1)}(s, t)$ are defined as follows
\bal
F^{(1)}(s) &=\frac{1}{2}\( \frac{\sqrt{z_s}}{16\pi^2} \ln \frac{\sqrt{z_s}-1}{\sqrt{z_s}+1} + \f{1}{8\pi^2}\)  \left(s^{2}-1\right), ~~~~~~z_x=1 - \f{4}x  ,
\\
G^{(1)}(s, t) &=\frac{1}{6} \( \frac{\sqrt{z_t}}{16\pi^2} \ln \frac{\sqrt{z_t}-1}{\sqrt{z_t}+1} + \f{1}{8\pi^2}\) \left(14-4 s-10 t+s t+2 t^{2}\right) .
\eal
Again, up on the twice subtraction, $b_1$ and $b_2$ do not enter the positivity bounds. $b_3$ and $b_4$ are related to the scale-independent $\mc{O}(p^4)$ low energy constants $\bar l_1$ and $\bar l_2$ (or the renormalized low energy constants $l_{1}^{r}(\mu)$ and $l_{2}^{r}(\mu)$) as follows
\bal
b_3 &= \f{1}{16\pi^2} \(\f13 \bar{l}_{1}+\f16\bar{l}_{2}-\frac{7}{12}\),~~~~~l_{1}^{r}(\mu)=\frac{1}{96 \pi^{2}}\left(\bar{l}_{1}+\ln \frac{M_{\pi}^{2}}{\mu^{2}}\right) ,
\\
b_4&=\f{1}{16\pi^2} \( \f16\bar{l}_{2}-\frac{5}{36}\),~~~~~l_{2}^{r}(\mu)=\frac{1}{48 \pi^{2}}\left(\bar{l}_{2}+\ln \frac{M_{\pi}^{2}}{\mu^{2}}\right) .
\eal
We shall formulate the bounds in terms of $\bar l_1$ and $\bar l_2$.

\begin{figure}
\begin{center}
\includegraphics[width=.45\linewidth]{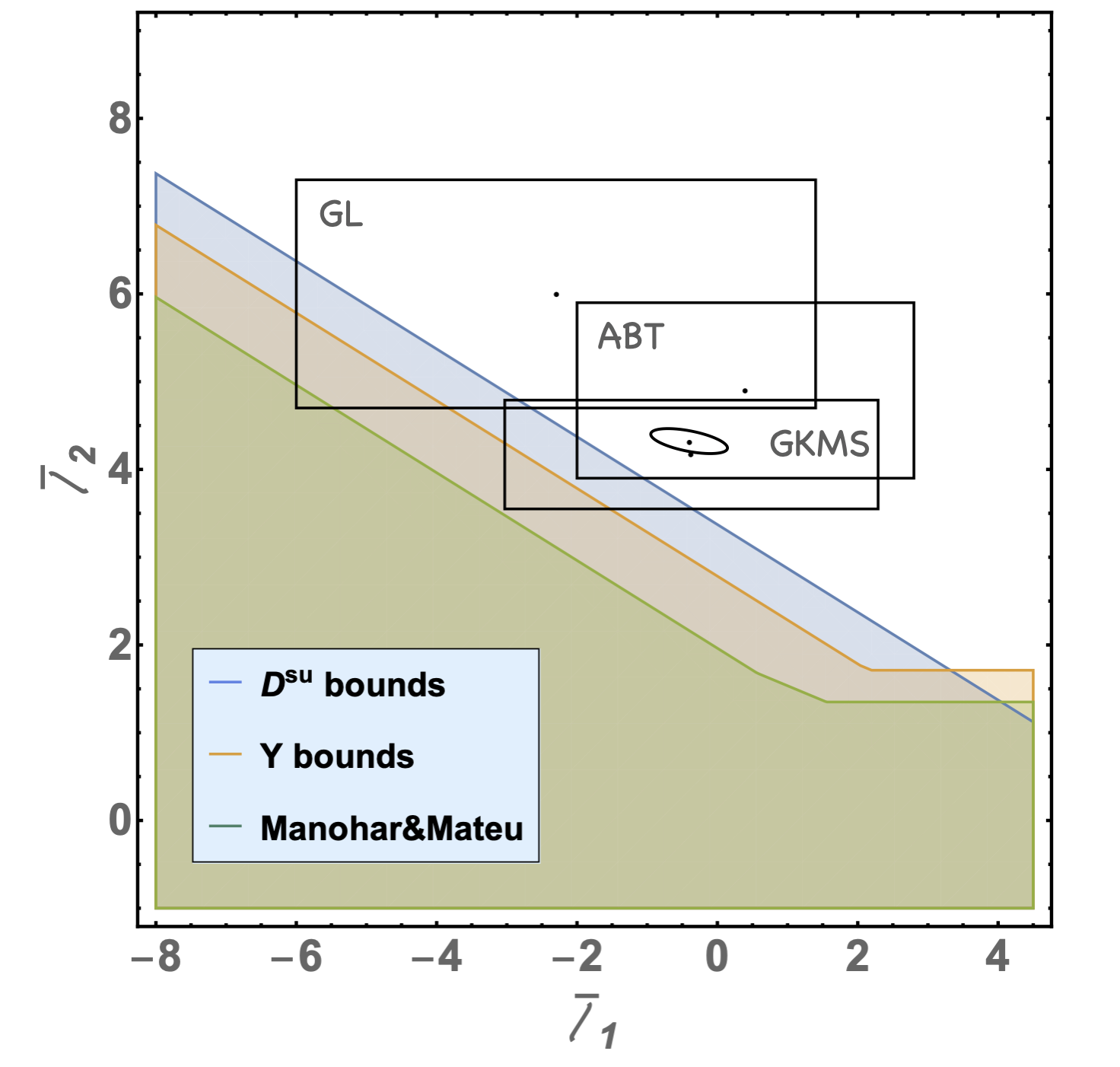}
\end{center}
\caption{Comparison of the $D^{\rm su}$ bound with the previous bounds on $\bar l_1$ and $\bar l_2$. The colored regions below the corresponding lines are ruled out by positivity. The Manohar\&Mateu bounds are obtained in \cite{Manohar:2008tc} by exploring the maximum analyticity of the amplitude in the Mandelstam plane for the $s$ derivative positivity bounds, and the $Y$ bounds are obtained in \cite{Wang:2020jxr} by the generalized bounds with both $s$ and $t$ derivatives that are also valid away from the forward limit. The ``GL'', ``ABT'', ``GKMS'' rectangle and the small ellipse are respectively the range of the fitted values of $\bar{l}_1$ and $\bar{l}_2$ from \cite{Gasser:1983yg}, \cite{Girlanda:1997ed}, \cite{Amoros:2000mc} and \cite{Colangelo:2001df}.}
\label{fig:chptfig}
\end{figure}

Since we have truncated the amplitude up to $\mc{O}(M_\pi^4/F_\pi^4)$ (or really $\mc{O}(M_\pi^4/\Lambda^4)=M_\pi^4/ (4\pi F_\pi)^4$), $b_{3,4}$ (and thus $\bar l_{1,2}$) only appear in the quadratic terms in the amplitude (cf.~\eref{Astupi}), so we only use bounds that at least contain one of $c_{m,n}$ with $2m+n\leq 4$. If a bound only contains $c_{m,n}$ with $2m+n> 4$, $\bar l_{1,2}$ will not appear in the bound, the bound being just a consistent relation without $\bar l_{1,2}$. Another observation is that up to $\mc{O}(M_\pi^4/F_\pi^4)$ only the term $M_\pi^4\left[b_{3} s^{2}+b_{4}(t-u)^{2}\right]/F_\pi^4$ in the amplitude enters the positivity bounds with dependence on the low energy constants. Since this term only contributes $c_{1,0}$ in the positivity bounds, we can use the bounds containing $c_{1,0}$ to constrain the low energy constants, while the other bounds can be used as consistency checks. The linear $PQ$ bounds do not contain anything different from the $Y$ bounds. The $c_{1,0}$ dependent bounds are the first nonlinear $PQ$ bound
\beq
c_{1,0} ~ c_{3,0} > c_{2,0}^2  \marrow  \bar{l}_1+2\bar{l}_2  > 3.85 ,
\eeq
and the first $D^{\rm su}$ bounds
\beq
c_{1,1}+\frac{3}{2}\sqrt{c_{1,0} ~ c_{2,0}} > 0 \marrow \bar{l}_1+2\bar{l}_2 > 6.74  .
\eeq
Clearly, the first $D^{\rm su}$ bound is stronger. We want to emphasize that since $M_\pi$ and $F_\pi$ factor out in these bounds, these bounds are independent of these parameters, that is, these are universal bounds on any SU(2) chiral perturbation theory, be it from quantum chromodynamics or from other UV models. See Fig.~\ref{fig:chptfig} for the comparison of the $D^{\rm su}$ bound with the previous bounds. We see that our $D^{\rm su}$ bound is stronger than the previous bounds.

The above bounds are obtained without using the $\epi\Lambda$ improved subtraction in \eref{Bdisrel}, where we estimate the low energy part (from $4$ to $(\epi\Lambda)^2$ since in this section we use the units where $M_\pi=1$) of the dispersive integral by using the EFT amplitude. In this language, the bound in Fig.~\ref{fig:chptfig} is the one with $\epi\Lambda=2$. If we choose $\epi \Lambda=2.2$, we would get $\bar{l}_1+2\bar{l}_2 > 7.23$, and  if we choose $\epi \Lambda=2.4$, we would get $\bar{l}_1+2\bar{l}_2 > 7.81$. How much subtraction one can use depends on how much errors introduced by using the EFT amplitude in the dispersive integral. A phenomenological analysis of the completion of errors and its application to hadron physics is left for future work.

\section{Summary}
\label{sec:sum}

The parameter space of the Wilson coefficients of an EFT is often vast. However, much of it may turn out to be some kind of ``uninhabitable swampland'', at least if we assume standard properties for the UV completion. By requiring the existence of a local analytic UV completion for the EFT, we can derive dispersion integral representations for the scattering amplitude, and use them to derive the positivity bounds to impose theoretical bounds on the vast parameter space. In this paper, we have supplied the tool box of positivity bounds with a few new sets of  bounds, again using the dispersion relation but now with finer properties of the Legendre polynomials (Gegenbauer polynomials in general dimensions) and the triple crossing symmetry of the amplitude. We have restricted to the simple case of single scalar fields for which crossing is straightforward.

While it is relatively straightforward to generalize the $PQ$ and $D^{\rm su}$ bounds to generic field theories with spin, the $D^{\rm stu}$ and $\bar{D}^{\rm stu}$ bounds, which make use of triple crossing symmetry, are more involved to implement, which will be discussed elsewhere \cite{moreBoundsSpin}.

 For a quick reference, we summarize the new positivity bounds that have been derived here:

\begin{itemize}

\item The nonlinear $PQ$ bounds
\beq
\mtP_{m,n}~\mtP_{m+2,n}> (\mtQ_{m+1,n})^2   ,
\eeq
where $P_{m,0} \= Q_{m,0} \= c_{m,0}$ and
\bal
\mtP_{m,n}& \= c_{m,n}+\frac{1}{(\epi\hat\Lambda)^2}\sum_{i=1}^{\lfloor\frac{n+1}{2}\rfloor}\mtM^{2i-1}_m
\mtP_{m+i-1,n+1-2i}-\sum_{j=1}^{\lfloor\frac{n}{2}\rfloor} \mtM^{2j}_m\mtQ_{m+j,n-2j}  ,
%> \<\frac{\mtL^{n}_\ell}{\mu^{2m+n-2}}\>
\\
\mtQ_{m,n}& \= c_{m,n}+(\epi\hat\Lambda)^2\sum_{i=1}^{\lfloor\frac{n+1}{2}\rfloor} \mtM^{2i-1}_m
\mtQ_{m+i,n+1-2i}-\sum_{j=1}^{\lfloor\frac{n}{2}\rfloor} \mtM^{2j}_m\mtP_{m+j,n-2j}  ,
%< \<\frac{\mtL^{n}_\ell}{\mu^{2m+n-2}}\>
\eal
with $(\epi\hat\Lambda)^2=(\epi\Lambda)^2-2m^2$ and $\mtM^i_m \= \left[{(m-1+i)!}/{((m-1)!i!)}+{(m+i)!}/{(m!i!)}\right]/2$.

\item The linear $PQ$ bounds
\bal
\mtP_{m,n} &> 0   ,
\\
\mtP_{m,n} &> (\epi\hat\Lambda)^{4k} \mtQ_{m+k,n}, ~~~k=0,1,2,...   .
\eal

\item  The $D^{\rm su}$ bounds
\bal
c_{m,2k}  + \sum_{i\geq 1} k_i c_{m+ i,2k-2i} & > S_{m,2k}(k)~c_{m+k,0}   ,
\\
c_{m,2k+1}  + \sum_{i\geq 1} k_i c_{m+ i,2k+1-2i}  & >  S_{m,2k+1}(k) \sqrt{c_{m+k,0}c_{m+k+1,0}}  ,
\eal
where $S_{m,n}(k)= \min_\ei D^{\rm su}_{m,n}(\ei,k)$, ${D}^{\rm su}_{m,n} (\eta,k) \equiv   \sum_{i\geq 0} k_i D_{m+ i,n-2i}(\eta)$ with $k_0=1$ and $D_{m,n}(\eta)$ is defined in \eref{Dk+} and \eref{Dk-}.

\item  The $D^{\rm stu}$ bounds
\bal
c_{m,2k}  + \sum_{i\geq 1} k_i c_{m+ i,2k-2i} & > U_{m,2k}(k) c_{m+k,0}  ,
\\
c_{m,2k+1}  + \sum_{i\geq 1} k_i c_{m+ i,2k+1-2i}  & >  U_{m,2k+1}(k) \sqrt{c_{m+k,0}c_{m+k+1,0}}  ,
\eal
where $U_{m,n}(k)=\max_{\ki}\min_\ei {D}^{\rm stu}_{m,n}(\eta,k,\ki)$, ${D}^{\rm stu}_{m,n}(\eta,k,\ki)=D^{\rm su}_{m,n}(\eta,k)+\sum_{j\geq 0}\ki_j \Gi_{m'+j,n'-2j}(\ei)$ and $\Gi(m,n)$ can be computed from \eref{GiDD} to \eref{Gihigh}.

\item The $\bar{D}^{\rm stu}$ bounds
\bal
c_{m,2k}  + \sum_{i\geq 1} k_i c_{m+ i,2k-2i} & < -T_{m,2k}(k)~c_{m+k,0}  ,
\\
c_{m,2k+1}  + \sum_{i\geq 1} k_i c_{m+ i,2k+1-2i}  &<  -T_{m,2k+1}(k) \sqrt{c_{m+k,0}c_{m+k+1,0}}   ,
\eal
where $T_{m,n}(k)={\rm max}_{\ki} {\rm min}_\ei  \bar{D}^{\rm stu}_{m,n}(\eta,k,\ki)$, $\bar{D}^{\rm stu}_{m,n}(\eta,k,\ki) =\! -D^{\rm su}_{m,n}(\eta,k) + \sum_{j\geq 0}\ki_j \Gi_{m'+j,n'-2j}(\ei)$ and $\ki_0>0$.

\end{itemize}

We have applied the new positivity bounds to weakly broken Galileon theories and found that in the case of the massive Galileon the new bounds constrain the cutoff of the theory to be parametrically close to the mass of the theory.  Similarly in the generic case of a weakly broken Galileon symmetry for other EFT operators, the symmetry breaking is forced to be at least of order unity. Thus, to be compatible with a standard local analytical UV completion, weakly broken Galileon theories are ruled out.
This does not preclude Galileon theories from playing a role in some UV completion, but necessitates that at least one of the standard assumptions, for instance locality, should be given up \cite{Dvali:2012zc,Keltner:2015xda}. It does however confirm that the apparent marginality of the massless Galileon, which appeared to be evaded by including a small breaking term \cite{deRham:2017imi} is in fact not marginal at all. \\

We have shown that the new bounds are complementary to each other and can in general significantly reduce the parameter space of allowed Wilson coefficients, improving the results of the previous $Y$ bounds. We have briefly considered chiral perturbation theory as an illustrative example of this. It is unlikely that we have exhausted the full implications of triple crossing symmetry in the bounds discussed, not of the specific details of the partial wave expansion, and it would be interested to further refine these arguments. We shall consider the extension of these bounds to massive particles with spin elsewhere \cite{moreBoundsSpin}, where the application of positivity bounds away from the forward limit is significantly more challenging \cite{deRham:2017zjm}.

\acknowledgments

We would like to thank Claudia de Rham, Yu-tin Huang, Scott Melville, Yu-Jia Wang and Cen Zhang for helpful discussions.
 The work of AJT  is supported by STFC grants ST/P000762/1 and ST/T000791/1.  AJT thanks the Royal Society for support at ICL through a Wolfson Research Merit Award.
SYZ acknowledges support from the starting grants from University of Science and Technology of China under grant No.~KY2030000089 and GG2030040375, and is also supported by National Natural Science Foundation of China under grant No.~11947301 and 12075233, and supported by the Fundamental Research Funds for the Central Universities under grant No.~WK2030000036.\\

\appendix

\section{Gegenbauer polynomials}
\label{sec:gegen}

The Gegenbauer polynomials are orthogonal polynomials defined via the generating function
\be
\label{GegenP}
\frac{1}{(1-2xt+t^2)^\ai} = \sum_{\ell=0}^\infty C^{(\ai)}_{\ell}(x) t^\ell    ,
\ee
which can be used to perform partial wave expansion in $D=2\ai+3$ dimensions when $D\geq 4$. $C^{(\frac{1}2)}_l(x) = P_l(x)$ is just the Legendre polynomial and $C^{(1)}_l(x) = U_l(x)$ is the Chebyshev polynomial of the second kind. Applying $n$-th $x$ derivatives on the both sides of Eq (\ref{GegenP})  gives
\be
\frac{(2t)^n \ai (\ai+1)...(\ai+n-1)}{(1-2xt+t^2)^{\ai+n}} = \sum_{\ell=0}^\infty \frac{\d^n}{\d x^n}C^{(\ai)}_{\ell}(x) t^\ell   ,
\ee
and expanding $t$ in the left hand side gives
\bal
\left. \frac{\d^n}{\d x^n}C^{(\ai)}_{\ell}(x)\right|_{x=1}&=\frac{ 2^{-2 \ai -n+1}\sqrt{\pi } \Gamma (\ell+n+2 \ai )}{ \Gamma (\ell-n+1) \Gamma (\ai )\Gamma \left(n+\ai +\frac{1}{2}\right)}  , ~~n=0,1,2,...  ,
\eal
which can shown to be positive. The Gegenbauer polynomials also satisfy the recurrence relation
\bal
C_{\ell}^{(\alpha)}(x) &=\frac{1}{\ell}\left[2 x(\ell+\alpha-1) C_{\ell-1}^{(\alpha)}(x)-(\ell+2 \alpha-2) C_{\ell-2}^{(\alpha)}(x)\right]   .
\eal
with $C_{0}^{(\alpha)}(x) =1$ and $C_{1}^{(\alpha)}(x) =2 \alpha x$.

\section{The $D=3$ case}
\label{sec:Dequto3}

When $D\leq 3$ (or $\ai\leq 0$), the Gegenbauer polynomials are usually not defined (see the generating function Eq (\ref{GegenP})). For $D<3$, there is no scattering angle, so it is not really meaningful to use the partial wave expansion. When $D=3$, Eq \ref{Cl1} blows up due to $\Gi(0)$ in the denominator. However, one can certainly decompose different angular momenta modes in $3$D. To this end, sometimes Gegenbauer polynomials with $\ai=0$ are defined as \cite{Gegen00}
\be
C^{(0)}_0(x)=1,~~~C^{(0)}_\ell(x) = \f2\ell T_\ell(x) ,~~~\ell>0   ,
\ee
where $T_\ell(x)$ is the Chebyshev polynomial of the first kind that can be defined as
\be
T_\ell(\cos\thi) = \cos \ell\thi   .
\ee
Directly using this expression to evaluate derivatives of $T_\ell(x)$ at $x=1$ leads to a $0/0$ indeterminate, but careful considerations give
\be
\lt.\f{\d^n}{\d x^n} T_\ell(x) \rt|_{x=1} =\prod^{n-1}_{k=0} \f{\ell^2-k^2}{2k+1} =   \f{\ell^2(\ell^2-1^2)...(\ell^2-(n-1)^2)}{1\cdot 3\cdot...\cdot(2n-1)} \geq 0   .
\ee
Combining this with partial wave unitarity, we still have
\be
\f{\d^n }{\d t^n}\im A(s,t) >0 , \quad  \quad s \geq 4 m^{2}, ~~ 0 \leq t<4 m^{2}       .
\ee
So the nonlinear $s$ derivative bounds, which have no explicit $D$ dependence, still hold. Furthermore, with appropriate normalization, the subtracted dispersion relation is given by
\be
B(s,t) = a(t) +  \sum_\ell \int \d \mu \[ \frac{s^2}{\mu-s}+\frac{u^2}{\mu-u}   \] \mu\ri_{\ell}(\mu) T_{\ell}\(1+\frac{2t}{\mu-4m^2}\)   .
\ee
Then following similar steps as in the generic case, we can explicitly check that the generic nonlinear $t$-derivative positivity bound \eref{upperlower} can be applied to the case of $D=3$ --- just evaluating $D$ in the equations at $D=3$.

\bibliographystyle{JHEP}
\bibliography{refs}

\end{document}